\RequirePackage{rotating}
\documentclass{aa}  
\usepackage{graphicx}
\usepackage{mathtools}
\usepackage[varg]{txfonts}
\usepackage{natbib}
\usepackage{subfig}
\usepackage{longtable}
\usepackage{bm}
\usepackage[colorinlistoftodos]{todonotes}
\usepackage{hyperref}
\usepackage{rotating}
\bibpunct[; ]{(}{)}{,}{a}{}{;}
\newcommand{\bd}{BD+46$^{\circ}$442}

\newcommand{\dd}{\mathrm{d}}
\newcommand{\iras}{IRAS19135+3937}
\begin{document}

	\title{A spatio-kinematic model for jets in post-AGB stars\thanks{Based on observations made with the Mercator Telescope, operated on the island of La Palma by the Flemish Community, at the Spanish Observatorio del Roque de los Muchachos of the Instituto de Astrofísica de Canarias.}\thanks{Reduced spectra are only available in electronic form at the CDS via anonymous ftp to cdsarc.u-strasbg.fr (130.79.128.5) or via http://cdsweb.u-strasbg.fr/cgi-bin/qcat?J/A+A/}
	}
	
	\author{D. Bollen
    	\inst{1,2}, D. Kamath \inst{2,3}, H. Van Winckel \inst{1}, O. De Marco \inst{2,3}
        }

	\institute{Instituut voor Sterrenkunde (IvS), KU Leuven,
              Celestijnenlaan 200D, B-3001 Leuven, Belgium\\
              \email{dylan.bollen@kuleuven.be}
              \and
              Department of Physics \& Astronomy, Macquarie University, 
              Sydney, NSW 2109, Australia
              \and
              Astronomy, Astrophysics and Astrophotonics Research Centre, 
              Macquarie University, Sydney, NSW 2109, Australia
              }

   \date{}
   \authorrunning{Bollen et al.}

 
   \abstract
  {}
  {
  We aim to determine the geometry, density gradient, and velocity structure of jets in post-asymptotic giant branch (post-AGB) binaries. }
  {Our high cadence time series of high-resolution optical spectra of jet-creating post-AGB binary systems provide us with a unique tomography of the jet. We determine the spatio-kinematic structure of the jets based on these data by fitting the synthetic spectral line profiles created by our model to the observed, orbital phase-resolved, H$\alpha$-line profiles of these systems. The fitting routine is provided with an initial spectrum and is allowed to test three configurations, derived from three specific jet launching models: a stellar jet launched by the star, an X-wind, and a disk wind configuration. We apply a Markov-chain Monte Carlo routine in order to fit our model to the observations. Our fitting code is tested on the post-AGB binary \iras.}
  {We find that a model using the stellar jet configuration gives a marginally better fit to our observations. The jet has a wide half-opening angle of about $76\,\degr$ and reaches velocities up to $870\,$km s$^{-1}$.}
  {Our methodology is successful in determining some parameters for jets in post-AGB binaries. The model for \iras\, includes a transparent, low density inner region (for a half-opening angle $<\,40\,\degr$). The source feeding the accretion disk around the companion is most likely the circumbinary disk. We will apply this jet fitting routine to other jet-creating post-AGB stars in order to provide a more complete description of these objects.}  

   \maketitle
%

\section{Introduction}
\label{sec:intro}
Astrophysical jets are frequently observed phenomena in the Universe, ranging from extremely energetic jets in active galactic nuclei \cite[AGNs,][]{blandford18} to non-relativistic stellar jets emerging from young stellar objects \cite[YSOs,][]{frank14} and planetary nebulae \citep[PNe,][]{livio99}. These jets are an important source of feedback to the interstellar medium and can influence the evolution of the launching engine and of the ambient medium \citep[and references therein]{soker16}.

Although a wide variety of jets are launched by distinct astrophysical objects, their general structure is very similar. In general, jets originate from a compact central object, they are two-sided, and have a certain degree of collimation, caused most likely by a magnetic field \citep{pino05}. Observations of YSOs clearly show two-sided or bipolar jets \citep{reipurth97, reipurth00, reipurth02, bally02, heathcote96, burrows96}. Observations of molecular outflows in YSOs also show two jet types: a wide-angle outflow and a highly collimated jet \citep{lee00}. The narrow, high-velocity jet propagates through the ambient medium and ends in a bow shock. These narrow jets are often accompanied by a slower, wide-angled outflow component which also carves a cavity in the ambient medium \citep{bally16, melnikov18}. 

Collimated outflows also exist in evolved objects such as PNe and have been extensively modelled (e.g., \citet[][Orion Source I]{greenhill98}, \citet[][DG Tauri]{bacciotti00}, \citet[][RW Aur, TH 28, and LkH$\alpha$ 321]{coffey04}, \citet[][He 3-1475]{contreras01}). In this contribution, we focus on low- and intermediate-mass, binary post-AGB stars recently discovered to have jets \citep{gorlova12,gorlova15, bollen17}. Our aim is to characterise these systems and determine the jet launching mechanism. 

Binary interactions during the AGB phase have a significant impact on the evolution of the binary system. When the binary orbit is small and the expanding AGB star grows larger than its Roche lobe, mass transfer will ensue and the orbital elements will evolve. If unstable mass transfer takes place, the binary will go through a common envelope phase with substantial orbital shrinkage \citep{ivanova13}, leaving behind a close binary system \citep{demarco09}. If the mass transfer is more stable, or if the initial orbital separation is large enough that Roche lobe overflow is not achieved, then the orbital separation might not decrease as much or even grow. A third possibility is an intermediate configuration with substantial mass transfer that however avoids a classical common envelope or delays it \citep{reichardt18} and may result in a post-AGB binary system with periods of 100 to 2000 days \citep{vanwinckel09}\footnote{If such an interaction takes place on the red giant branch, a post-red giant branch system will result instead \citep[][]{kamath16}}.

Binary post-AGB systems display a distinctive near-infrared (near-IR) excess in their spectral energy distribution (SED) that has been shown to be a signature of a stable, Keplerian circumbinary disk \citep{vanwinckel03, bujarrabal13a, kamath14a, kamath15a}. This implies a connection between the binary interaction and the formation of the circumbinary disk.

The post-AGB binary system therefore comprises a circumbinary disk, an evolved post-AGB star (the primary component), and a companion, likely to be a main sequence star \citep{oomen18}. Using interferometric image reconstruction techniques, signatures of an additional component have been detected in the post-AGB binary system IRAS08544-4431 \citep{hillen16a, kluska18}, that are interpreted as an accretion disk around the companion, possibly fed by the primary or by reaccretion of circumbinary material.

High resolution, radial velocity (RV) monitoring of several post-AGB stars by \cite{vanwinckel09} has detected variable absorption features interpreted as jets emanating from the companion \citep{gorlova12, gorlova14, gorlova15, bollen17}. During superior conjunction, when the companion is closest to the observer and the evolved star is behind, continuum photons from the primary are scattered out of the line-of-sight by the H-atoms in the jet, resulting in a P-Cygni line profile during these orbital phases. Since the light from the primary shines through different parts of the jet during the orbital motion of the binary system, these time-series provide us with a tomography of the jet that may uniquely constrain the spatial, velocity, and density structure of the jet.

\cite{bollen17} fitted the time-series spectra of the post-AGB binary \bd\ with a simple conical jet model and determined a rather wide jet opening angle ($>$100\degr), where a low density fast central outflow is enclosed in a denser and slower outer layer.

We have now observed that a large fraction of well-monitored post-AGB binaries show the presence of similar features, indicating that jets are commonly found in these systems. In this study, we present a refined technique to determine improved parameters such as outflow velocities and density scales, which are crucial to establish the jet outflow momenta and determine mass-accretion rates as a function of orbital parameters. We use the MCMC technique to fit the observed H$\alpha$-line profile variations. The strength of this work and its application of MCMC is that we now have an automated model-fitting code, which provides more information on the complete kinematic, geometrical, and density structure of these jets. The ultimate aim of our study is to apply our fitting routine to all post-AGB binary systems with jets, so as to investigate the observed diversity of the jets, understand the jet launching mechanism, and the dynamics of the interaction that leads to the formation of post-AGB binaries.

This paper is structured as follows: in Sect. \ref{sec:model}, we present the jet model that we implemented in this code. In Sect. \ref{sec:MCMC}, we elaborate on our model-fitting routine. In Sect. \ref{sec:IRAS}, we will illustrate our jet model applied to \iras, a post-AGB binary system \citep{gorlova15}, which is presented in Sect. \ref{sec:IRAS}. We discuss the results of our fitting routine in Sect. \ref{sec:spatiomod}. We finish with our conclusions in Sect. \ref{sec:conclusions}.


\section{The post-AGB jet model}
\label{sec:model}

\begin{figure}[h!]
\centering
\includegraphics[width=.5\textwidth]{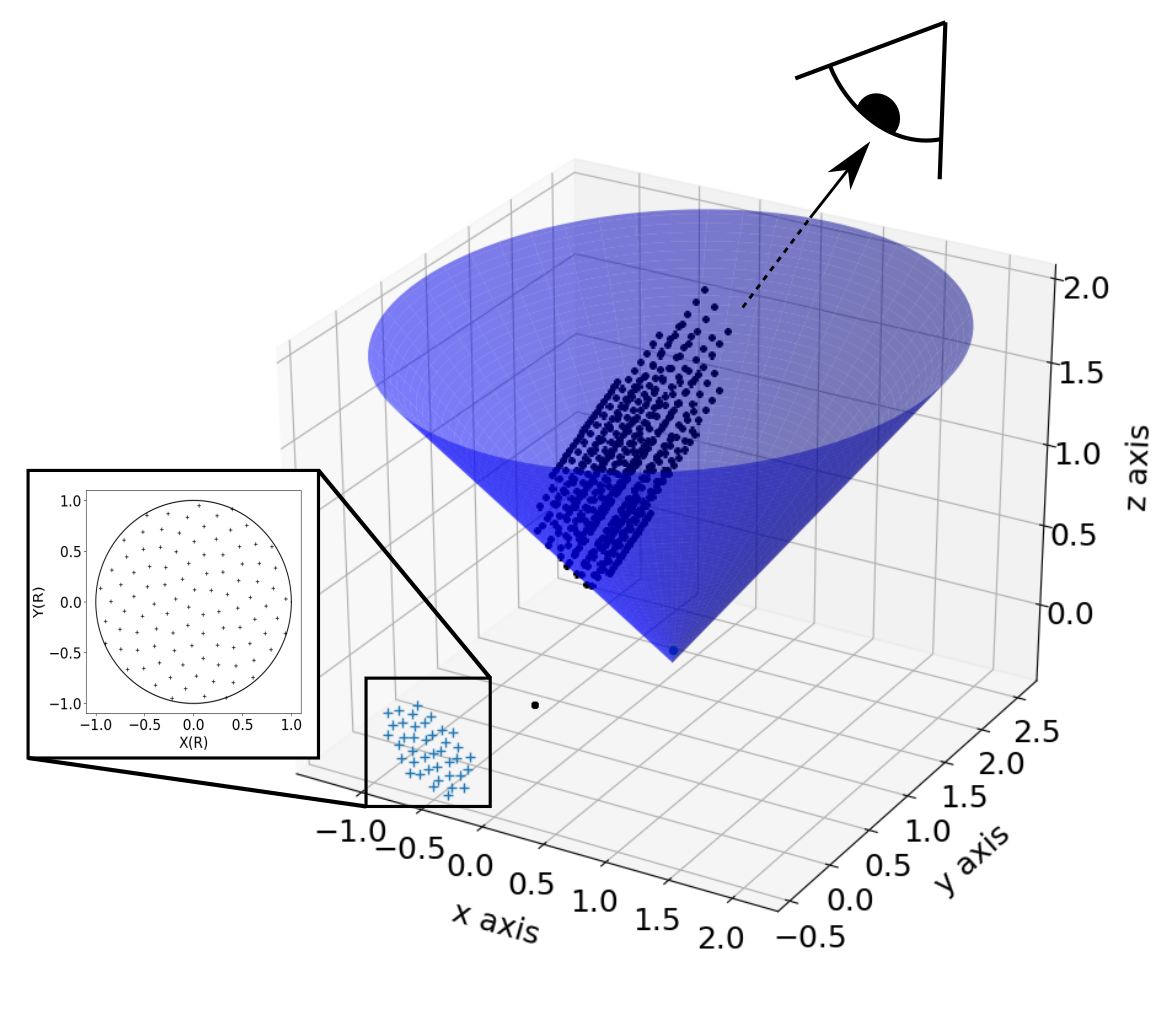}
\caption{Mesh grid of the primary star component and the rays going through the jet. The primary component is sampled as a Fibonacci grid with 200 grid points. The absorption of the continuum light by the H-atoms in the jet is determined for each ray along the line-of-sight that departs from these points on the primary's surface.}\label{fig:grid}
\end{figure}

The central stellar binary system in our model consists of an evolved star (the post-AGB star) and a companion (the low-mass main sequence star). We model the post-AGB star as a uniform disk that is always perpendicular to the line-of-sight, because the evolved star will be projected as a disk along the line-of-sight towards the observer. We divide this disk up into a Fibonacci grid, whose benefit is that each grid point is a surface with equal area, as can be seen in Fig. \ref{fig:grid}. We follow the light of the evolved star commencing from these grid points, along the line-of-sight towards the observer, to calculate the absorption by the jet.

The luminosity of the companion star in these systems is often negligible relative to the primary's luminosity. Hence, its flux contribution is too small to be observed in the spectra. For this reason, we consider the total flux to be from the primary and ignore the flux contribution of the companion star in our model.

In post-AGB binary systems, jets are launched from the accretion disk around the companion. The jets travel in opposite directions, perpendicular to the orbital plane of the binary system. During each orbital phase, we determine which rays from the evolved star pass through the jet cones in our model. We then follow these rays through the jet. Each ray through the jet is sampled by a number of grid points between the point at which the ray enters the jet and the point at which it exits. At these grid points, we calculate the local density, velocity, and velocity dispersion, from which the optical depth due to line scattering can be calculated over the whole ray and for all frequencies. 

In the following subsections, we will describe the geometrical structures of the jet that we implement in our model. We will then describe how we recreate the absorption feature in our model that is caused by the jet occultation by giving a detailed description of the radiative transfer of the AGB star light from the evolved star through the jet. 

\subsection{Jet geometry}
We aim to determine the structure of jets launched in post-AGB binary systems. We choose a jet geometry based on jet launching mechanisms originally developed for YSOs, because of the similarity between the observational data between young stars and our evolved systems. Over the past decades, several jet launching theories have been developed in order to explain YSO jets. The most widely accepted ones are those of \citet[][Model A]{matt05}, the magneto-centrifugally accelerated disk wind model by \citet[][Model B]{blandford82}, and the X-wind model by \citet[][Model C]{shu94}.

These jet models are applied by \citet{kurosawa06, kurosawa11}, \citet{weigelt11}, and \citet{tambovtseva14} to determine the structure and physical properties of jets in YSOs. In these studies the authors determined the density and velocity structure of jets in classical T Tauri stars (CTTSs) by generating synthetic hydrogen and helium spectra and fitting those to their observations. Since these studies focused on the modeling of single star systems, the hydrogen and helium lines remain unchanged (neglecting changes due to changing accretion rates). Our work has extra complexity, which lies in the binary nature of the systems. This results in spectral lines that vary over the orbital period. We account for this by implementing the binary motion in our model, such that we can calculate the absorption caused by the jet during each orbital phase of the binary system.

In order to find which model best represents our diverse sample of post-AGB binaries, we implement all three and carry out fits to our data. Below, we give a detailed description of the velocity and density structure for each of these jet configurations.
\begin{figure*}[h!]
\captionsetup{width=1\textwidth}
\centering
\includegraphics[width=1\textwidth]{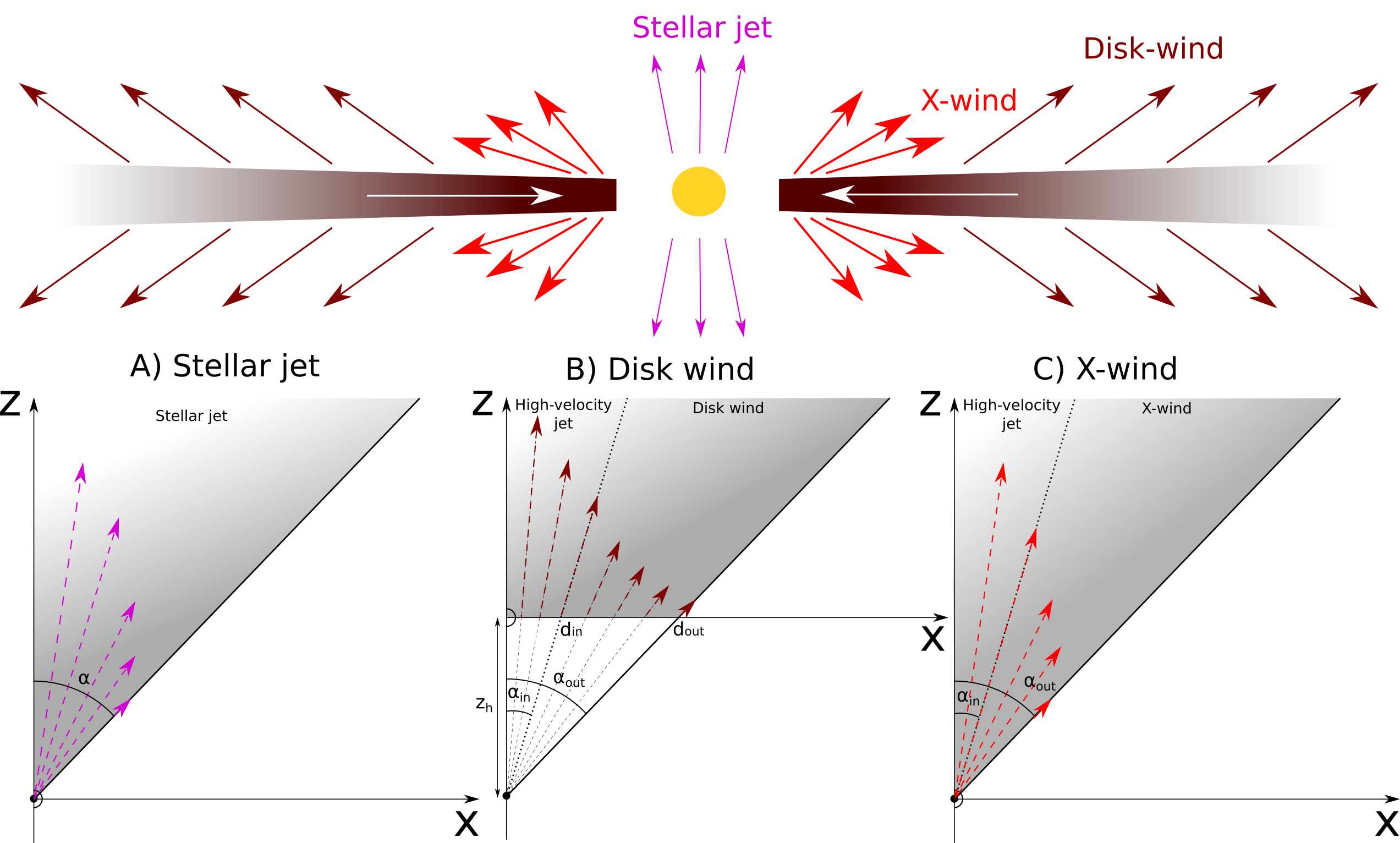}
\caption{\textit{Upper:} A schematic sketch of the jet model, showing our geometric implementation of the three models. \textit{Lower:} The launching region and velocity field of the jet for the stellar jet (A), the disk wind (B), and the X-wind (C). Unlike the stellar jet and X-wind, the launching region for the disk wind is the region between the inner-disk radius $d_\mathrm{in}$ and the outer disk radius $d_\mathrm{out}$. This configuration is based on the disk wind configuration applied by \cite{kurosawa06}.}\label{fig:schematicjet}
\end{figure*}

\subsubsection{Model A: ``stellar jet" configuration}
\label{sssection:stellarjet}
The first configuration is similar to that applied by \cite{kurosawa06}, who modeled the H$\alpha$ emission in CTTSs. They based this on the jet model by \cite{matt05}. According to \citet{matt05} the star, that is surrounded by an accretion disk, shows that the spin-down torque of the star is most likely due to the stellar wind, travelling along the open field lines of the star. The stellar magnetic field is strong enough to truncate the accretion disk, which causes disk material to be channelled along the magnetic field lines, part accreting onto the star and part being ejected. Thus, the ejected matter originates from the disk, but is ejected along the open stellar magnetic field lines that are anchored on the star and therefore we call this mode the ``stellar jet" model. The magnetic field lines of the companion star in our systems are assumed to be similar to those of YSOs\footnote{Even if our companions are old and likely more magnetically inactive, accretion during the interaction phases that created the post-AGB star may have rejuvenated them as can be inferred by activity on related binaries \citep{montez15}.}.

In our model, we assume that the jet travels along the open magnetic field lines of the companion star. Hence, the matter in the jet is ejected radially away from the star, as can be seen in panel A of Fig. \ref{fig:schematicjet}. We implement a latitudinally dependent velocity law for the matter inside the jet, as done by \cite{thomas13}. The first one is:
\begin{equation}
\vec{v} = \left[v_0+(v_\alpha-v_0)\left(\frac{\theta}{\alpha}\right)^2\right]\mathbf{\hat{r}} ,\label{eq:vtheta}
\end{equation}
where $\theta$ is the polar latitudinal coordinate, $\alpha$ the half-opening angle of the jet, $v_\alpha$ is the velocity at the jet edges and $v_0$ is the velocity along the jet axis. Hence, the jet velocities are independent of the radial component. 

Just like the jet velocity, the density at each position in the jet has a latitudinal dependency, where the jet has a low density along the jet axis and reaches the highest densities along the jet edges. The density decreases as a function of jet height $z$ as follows:
\begin{equation}
\rho(\theta, \mathrm{z}) \propto \frac{\theta^p}{z^2},    \label{eq:rho_stellar}
\end{equation}
By choosing different values for $p$, the density dispersion in the jet can be increased or decreased in our model. The range of adopted values for the factor $p$ is given in Table \ref{tab:pval}.


\subsubsection{Model B: ``disk-wind" configuration with inner jet}
\label{ssec:diskwindmodel}
\begin{table}[t!]
\begin{center}   
  \caption{Different values of the $p$ and $p_\mathrm{in}$/$p_\mathrm{out}$ parameters.}\label{tab:pval}
\begin{tabular}{cccc} \hline \hline
$p$ & $p_\mathrm{in}$ & $p_\mathrm{out}$ & Model\\
\hline
2 & & & A\\
4 & & & A\\
8 & & & A\\
 & 8 & 2 & B, C\\
 & 8 & -2 & B, C\\
 & 4 & 2 & B, C\\
 & 4 & -2 & B, C\\
 & 2 & -2 & B, C\\
 & 0 & 4 & B, C\\
\hline
\end{tabular}
\end{center}
\end{table}
The magneto-centrifugal disk wind model, first introduced by \cite{blandford82}, is suggested to be the main launching mechanism for jets from accretion disks around stellar objects, such as black holes and young stars \citep{pudritz83}. In this model, the magnetic field lines are anchored to the accretion disk and inclined with respect to the rotation axis of the accretion disk by at least $30\degr$. The matter in the disk below and above the mid-plane will be accelerated due to the magneto-centrifugal force and follow these magnetic field lines like beads on a wire.

In our model, we implement this disk wind as our second jet configuration. The launching region of the jet in this configuration is the circum-companion accretion disk. The inner and outer disk radii are determined from the inner and outer jet angles in the model, and have set limits: the inner disk radius must be larger than twice the radius of the companion star and the outer disk radius must be smaller than the Roche lobe. Since we do not know the exact size of the companion star, we assume a radius given by the stellar mass-radius relation by \cite{demircan91}:
\begin{equation}
R_\star = 1.01\,R_\odot  \left(\frac{M_\star}{M_\odot}\right)^{0.724}.
\end{equation}
To determine the Roche radius of the companion, we use the approximation by \cite{eggleton83}:
\begin{equation}
r_L = \frac{0.49\,q^{2/3}}{0.6\,q^{2/3} + \ln\,(1+q^{1/3})},
\end{equation}
Where $q=M_1/M_2$ is the mass ratio of the binary system.
The jet configuration has two outflowing components: the inner jet component and the disk wind component. The inner jet component has a similar velocity law as the stellar jet discussed in Section \ref{sssection:stellarjet}. The disk wind originates from the circum-companion disk, in which the particles are ejected latitudinally from the region between the inner and outer disk edges at angles between $\alpha_\mathrm{in}$ and $\alpha_\mathrm{out}$, as shown in panel B of Fig. \ref{fig:schematicjet}. The disk wind velocity is related to the Keplerian velocity of the disk. This means that the disk wind velocity of the particles ejected from a certain radius in the disk equals the Keplerian velocity at that position in the disk, multiplied with a scaling factor $c_v$. This scaling factor is a free parameter in our model and has a value between 0 and 1. The angle-dependent velocity law is defined as follows:
\begin{eqnarray}
v = \left\{
    \begin{array}{ll}
        c_v\cdot\sqrt{\frac{\mathrm{G}M_2}{d}} = v_\mathrm{\alpha, out} \sqrt{\frac{\tan\alpha_\mathrm{out}}{\tan\theta}}\, ;  & \mbox{if } \theta > \alpha_\mathrm{in} \\ \\
       \big[v_0+(v_\mathrm{\alpha,in}-v_0)\left(\frac{\theta}{\alpha_\mathrm{in}}\right)^2\big]\, ; & \mbox{if } \theta < \alpha_\mathrm{in}. \label{eq:vtheta_diskwind}
    \end{array}
\right.
\end{eqnarray}
with $c_v$ the scaling factor, $d$ the radial distance between the launch site in the disk and the centre of the companion star, $M_2$ the mass of the companion, and $v_\mathrm{\alpha,in} = c_v\cdot\sqrt{GM_2/d_\mathrm{in}}$ the Keplerian velocity at the inner disk rim $d_\mathrm{in}$. 

The density profile in the jet for this system is determined as follows:

\begin{eqnarray}
\rho(\theta, \mathrm{z}) \propto \left\{
    \begin{array}{ll}
        \left(\frac{\theta}{\alpha_\mathrm{in}}\right)^{p_\mathrm{out}}\,\mathrm{z}^{-2}\,; & 
        \mbox{if } \theta > \alpha_\mathrm{in} \\ \\
       \left(\frac{\theta}{\alpha_\mathrm{in}}\right)^{p_\mathrm{in}}\,\mathrm{z}^{-2}\, ; & 
       \mbox{if } \theta < \alpha_\mathrm{in}, \label{eq:rho_xwind}
    \end{array}
\right.
\end{eqnarray}


\subsubsection{Model C: ``X-wind" configuration with inner jet}
\label{ssec:Xwindmodel}
The X-wind theory for YSOs was introduced by \cite{shu94} for a scenario where the interaction between the magnetosphere of rapidly rotating YSOs and the surrounding accretion disks can give rise to the formation of so-called X-winds. In this model, the central star is surrounded by an accretion disk that extends inwards towards the star, down to an inner disk radius of about twice the stellar radius of the central star. The stellar magnetic field is not able to penetrate the accretion disk smoothly, however. This causes the magnetic field to be squeezed together in the inner disk region, named the X-region. A fraction of the inflowing matter in the disk will be launched from the equatorial plane by the magnetic field in the X-region, thus creating an X-shaped outflow. This wind efficiently carries away angular momentum from the disk. In other words, this outflow originates from the inner disk edge, where the disk's magnetic field and the co-rotating stellar field meet. A more detailed description of the X-wind can be found in \citet{shu97}. \cite{shang07} pointed out that the disk wind and the X-wind are intrinsically of the same nature: both are magneto-centrifugally accelerated winds, with the only difference being the truncation of the magnetic field in the accretion disk. Where the magnetic field of a disk wind can penetrate through the whole accretion disk, the field of an X-wind will only truncate the inner disk region.

In our model, the X-wind configuration comprises two main jet components, each with a distinct velocity profile. The true launching region of the X-wind is about twice the stellar radius of the main sequence star. As this is relatively small compared to the scale of the binary system, we consider the launching point in our model to be located at the position of the main sequence star, similar as the stellar jet model. The velocity structure in this model is defined as follows:
\begin{eqnarray}
\vec{v} = \left\{
    \begin{array}{ll}
        \Big[v_{\alpha\mathrm{, out}} + (v_\mathrm{M} - v_{\alpha\mathrm{, out}})\cdot \cos{\left(\frac{\pi}{2}\frac{\theta}{\alpha_\mathrm{out}}\right)} \Big]\,\hat{r}\, ; & \mbox{if } \theta > \alpha_\mathrm{in} \\ \\
        \Big[ v_{\alpha\mathrm{, out}} + (v_\mathrm{M} - v_{\alpha\mathrm{, out}})\cdot \cos{\left(\frac{\pi}{2}\frac{\theta}{\alpha_\mathrm{out}}\right)}  \\ \\
       \hspace{1cm}+  \left(v_0 - v_\mathrm{M} \right)\cdot \cos{\left(\frac{\pi}{2}\frac{\theta}{\alpha_\mathrm{in}}\right)} \Big]\,\hat{r}\, ; & \mbox{if } \theta < \alpha_\mathrm{in} \label{eq:vtheta_nested_cos}
    \end{array}
\right.
\end{eqnarray}
where $\alpha_\mathrm{out}$ and $\alpha_\mathrm{in}$ are the outer and inner jet angles at the boundary between the two velocity laws, respectively. The variables $v_\mathrm{\alpha, out}$ and $v_\mathrm{\alpha, in}$ are the radial jet velocities at the outer jet angle and inner boundary angle, respectively. Finally, $v_\mathrm{M}$ is defined as:

\begin{equation}
v_\mathrm{M} = \frac{v_\mathrm{\alpha, in} - v_\mathrm{\alpha, out}}{\cos\left(\frac{\pi}{2}\frac{\alpha_\mathrm{in}}{\alpha_\mathrm{out}}\right)} + v_\mathrm{\alpha, out}.
\end{equation}
This model is based on a velocity profile similar to that used by \cite{federrath14} who used this nested velocity profile in order to model the momentum transfer and feedback by jets and outflows launched from protostellar disks. 

The density profile in this model is equal to the density profile of the disk-wind model, given in Eq. \ref{eq:rho_xwind}.

With hindsight, it would have been best to leave a cavity in both the disk wind and X-wind models, as the original theoretical models in fact prescribe. However, the density law that we use implies a low density outflow close to the centre for these two models, which results in a central outflow.

\subsection{Radiative transfer through the jet}
\label{ssec:radiativetransfer}

To model the amount of absorption by the jet, we determine the loss of intensity of the photospheric contribution of the primary due to scattering of continuum photons from the primary by the hydrogen atoms in the jet. This causes the observed absorption feature in the H$\alpha$-profiles of the spectra during superior conjunction. Since the gas in the jet has high outflow velocities, the absorption by the jet lobe that is pointed towards us will be mainly blue-shifted. The jet emission is assumed negligible and we do not include this in our model. Hence, the specific intensity along a ray aligned with the line-of-sight becomes:

\begin{eqnarray}
I_\nu(n, s) =& &I_\nu^0(n) \,\mathrm{e}^{-\tau_\nu(n, s)},
\end{eqnarray}

\noindent where $I_\nu^0$ is the initial intensity, $\tau_\nu$ is the optical depth at position $s$ along the $n$-th line-of-sight from the primary component. The photospheric intensity along each line-of-sight from the primary is equally weighted. Since there is a large velocity gradient along the line-of-sight, we implement the Sobolev approximation to determine the optical depth in our medium. Hence, the optical depth will be proportional to:

\begin{equation}
\tau_\nu(n,s) \propto \frac{\chi(n,s)\,\Delta s}{\left| \dd v_s/\dd s \right|},
\end{equation}

\noindent where $\Delta s$ is the length of the line element at position $s$, $v_s = \vec{v}\cdot\vec{\hat{n}}$ is the projected velocity along the line-of-sight, and $\left| \dd v_s/\dd s \right|$ is the velocity gradient at position $s$. The optical depth is determined for each time step along each line-of-sight from the primary component towards the observer. 
The extinction coefficient $\chi(n,s)$ is

\begin{eqnarray}
\chi(n,s) = \frac{\pi\,e^2}{m_e\,c}f_\mathrm{lu}\left(n_\mathrm{l} - \frac{g_\mathrm{l}}{g_\mathrm{u}}n_\mathrm{u}\right), \label{eq:extcoeff}
\end{eqnarray}

\noindent where $m_e$ is the electron mass, $f_\mathrm{lu}$ is the oscillator strength, and where $l$ and $u$ denote the lower and upper levels of the transition. We assume hydrogen to be mainly neutral (HI), i.e. $n_\mathrm{HI}\approx n_\mathrm{H}$. According to the Boltzmann equation,

\begin{eqnarray}
\frac{n_i}{n_\mathrm{HI}} \approx \frac{n_i}{n_1} = \frac{g_i}{g_\mathrm{1}}\exp\left(-\frac{h\,\nu}{k_B\,T}\right), \label{eq:boltzmann}
\end{eqnarray}

\noindent where $k_b$ is the Boltzmann constant, $g$ is the statistical weight that takes into account the degeneracy of the energy states, $i \in \{l,u\}$, and where we assume that most of the hydrogen is in the ground state ($n_\mathrm{HI} \approx n_1$). Since the jet consists mainly of hydrogen and helium, we can write the mass density as

\begin{eqnarray}
\rho &\approx & n_\mathrm{H}\,m_\mathrm{H} + n_\mathrm{He}\,m_\mathrm{He} = n_\mathrm{H}\,m_\mathrm{H} + Y_\mathrm{He}\,4\,m_\mathrm{H}\nonumber \\
&=& n_\mathrm{H}\,m_\mathrm{H}\,(1+4\,Y_\mathrm{He}), \label{eq:massdensity}
\end{eqnarray}

\noindent where $n_\mathrm{H}$ is the hydrogen number density, $m_\mathrm{H}$ is the mass of a hydrogen atom, $n_\mathrm{He}$ is the helium number density, $m_\mathrm{He}$ is the mass of a helium atom, and $Y_\mathrm{He}=n_\mathrm{He}/n_\mathrm{H}$. By assuming the jet to be isothermal and in local thermodynamical equilibrium, we can use Eqs.~(\ref{eq:boltzmann}) and (\ref{eq:massdensity}) in Eq.~(\ref{eq:extcoeff}) to obtain

\begin{eqnarray}
\chi(n,s) \propto \rho(n,s).
\end{eqnarray} 

\noindent From this follows that

\begin{eqnarray}
\tau_\nu(n,s) \propto \frac{\rho(n,s)\,\Delta s}{\left| \dd v_s/\dd s \right|}.
\end{eqnarray} 

\noindent The resulting intensity will be the contribution of the intensity passing through the jet along each line-of-sight:

\begin{eqnarray}
I_\nu =\sum_n I_\nu^0(n) \,\prod_s \, \exp\left(-\frac{c\,\rho(n,s)\,\Delta s}{\left|\dd v_s/\dd s\right|}\right),
\end{eqnarray}

\noindent where $c$ is a scaling factor that is multiplied to the optical depth, such that the resulting intensity in our model can be scaled to the observed intensity in our spectra. The reason for the implementation of the scaling factor $c$ is that this method does not allow us to determine the absolute jet density\
In the following section we explain how we implement the fitting of each model to the observations so as to decide the best jet model.
	
\section{MCMC-fitting routine}
\label{sec:MCMC}
\begin{table}[h!]
\begin{center}
\caption{Boundary conditions and best-fitting values for our model parameters.}
\label{tab:boundaries}
\begin{tabular}{l cc}\\
\hline
\hline 
Model parameter & Grid range & Best-fitting value \\
\hline
$i$ ($\degr$) & $50-80$ & 78.8\\
$\theta_\mathrm{out}$ ($\degr$)  & $\theta_\mathrm{in}-80$ & 75.9\\
$\theta_\mathrm{in}$ ($\degr$) & $0-\theta_\mathrm{out}$ & \\
$c$  & $0-7.5$ & 1.94\\
$v_0$ (km s$^{-1}$) &  $400-1400$ & 1210  \\
$v_\mathrm{in}$ (km s$^{-1}$) & $v_\mathrm{out}-v_0$ & \\
$v_\mathrm{out}$ (km s$^{-1}$) & $0-150$ & 11\\
$c_v$ & $0.01-1$ & \\
$R_\mathrm{prim}$ $(\mathrm{a_{prim}})$ & 0.2-0.7  & 0.66\\
\hline
\end{tabular}
\end{center}
\tablefoot{See Sect. \ref{sec:MCMC} for further details on the individual parameters.}
\end{table}
In order to fit our model to the data, we use a Markov chain Monte Carlo (MCMC) routine. This was done with the \texttt{emcee} package \citep{foreman13}, a pure-Python implementation of Goodman \& Weare's ``Affine MCMC ensemble sampler" \citep{goodman10}.

The free parameters of this fitting routine in our three jet configurations are: inclination angle of the binary orbit, $i$, jet half-opening angle, $\theta_\mathrm{out}$, scaling factor for the optical depth, $c$, jet velocity along the jet axis, $v_0$, jet velocity at the edges, $v_\mathrm{out}$, and the radius of the primary, $R_\mathrm{prim}$ (in units of the semi-major axis $a_\mathrm{prim}$ of the primary). Depending on the adopted model, there are additional free parameters: the inner jet angle at the boundary between two velocity regions in the jet ($\theta_\mathrm{in}$ for model B \& C), the velocity at this boundary ($v_\mathrm{in}$ for model B), and the disk-wind scaling factor for the jet velocity ($c_\mathrm{v}$ for model C). 

Before starting the MCMC routine, we choose 128 initial combinations of the free parameters, where each combination is referred to as a ``walker". These walkers are given a random initial distribution between predetermined boundaries. Some additional constraints are imposed on the walkers, i.e., the outflow velocity in the jet decreases for increasing polar angle, $v_\mathrm{out} < v_\mathrm{in} < v_0$. Secondly, the primary radius $R_\mathrm{prim}$ is kept smaller than 70\% of the Roche radius of the primary component. This constraint is derived from the observations of our test object (see Section \ref{sec:subIRAS} for more details).
The full list of model parameters and corresponding constraints is given in Table \ref{tab:boundaries}. 

The walkers will start exploring parameter space from this initial position during the MCMC routine. The acceptance or rejection of a new position during each iteration depends on the log-likelihood of that new position. The log-likelihood is given by:
\begin{equation}
\ln\mathcal{L} = -\frac{1}{2}\, \sum_i\left[\frac{(y_i - m_i)^2}{\sigma_{\mathrm{tot,}\,i}^2} + \ln\left(2\,\pi\,\sigma_{\mathrm{tot,}\,i}^2\right) \right],\label{eq:logl}
\end{equation}
with $y_i$ the data, $m_i$ the model, and $\sigma_{\mathrm{tot,}\,i} = \sigma_{\mathrm{S/R,}\,i} + \sigma_{\mathrm{bs,}\,i}$ the error on the data. The total uncertainty $\sigma_{\mathrm{tot,}\,i}$ in our data is the sum of the error on the spectra determined from the signal-to-noise, $\sigma_{\mathrm{S/R,}\,i}$, and the uncertainty that comes from our determination of the background spectra $\sigma_{\mathrm{bs,}\,i}$ (see Section~\ref{sec:inputspec}). The sum $i$ is over all sampled wavelength points of our spectra. We run the MCMC chains until the walkers converge to a certain value for each parameter. The best-fitting parameters for each MCMC routine is chosen by fitting a Gaussian profile to the posterior probability distribution, which yields a mean and standard deviation for each parameter. We only use the iterations following the burn-in phase to determine the mean and standard deviations from the posterior distribution.


\section{Using observations of a post-AGB star with jets to constrain our jet model}\label{sec:IRAS}
\subsection{The post-AGB binary \iras}\label{sec:subIRAS}
\begin{figure}[t!]
\centering
\includegraphics[width=.5\textwidth]{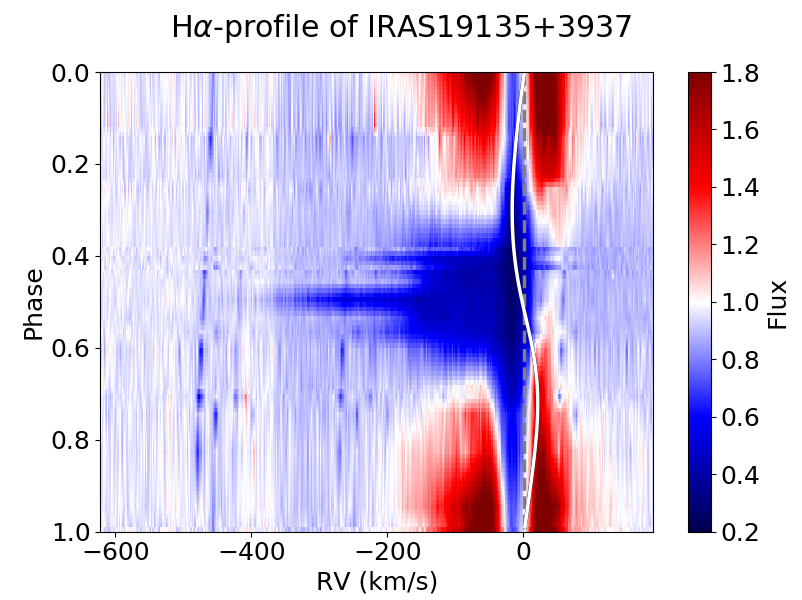}
\caption{Interpolated dynamic spectra for the H$\alpha$-line of IRAS19135+3937, based on the 22 spectra (see Sect.\ref{sec:subIRAS}). The dashed grey line indicates the systemic velocity ($\gamma$) of the binary system and the solid line indicates the RV of the post-AGB star.}\label{fig:dynspec}
\end{figure} 
\begin{table}[h!]
\begin{center}
\caption{Spectroscopic orbital solutions of the primary component of \iras\, \citep{oomen18}.}
\label{tab:orbpar}
\begin{tabular}{l cc}\\
\hline
\hline 
Parameter & Value & $\sigma$ \\
\hline
$P$ (d) & 126.97& 0.08\\
$T_0$ (BJD) & 2454997.7 & 1.0\\
$e$ & 0.13 & 0.03\\
$\omega$ ($^{\circ}$) & 66.0 & 4.4\\
$\gamma$ (km s$^{-1}$) &  31.7 & 0.3  \\
$K_\mathrm{prim}$ (km s$^{-1}$) & 18.0 & 0.6\\
$a_\mathrm{prim}\sin i$ $(\mathrm{AU})$ & 0.209 & 0.008\\
$f(\mathrm{m})$ $(M_\odot)$ & 0.075 & 0.008\\
\hline
\end{tabular}
\end{center}
\tablefoot{The tabulated orbital parameters are: orbital period $P$, time of periastron $T_0$, eccentricity $e$, argument of periastron $\omega$, systemic velocity $\gamma$, radial velocity of the primary $K_\mathrm{prim}$, projected semi-major axis of the primary $a_\mathrm{prim}\sin i$, and mass function $f(\mathrm{m})$.}
\end{table}
\begin{figure*}
\centering
  \includegraphics[width =1.\textwidth]{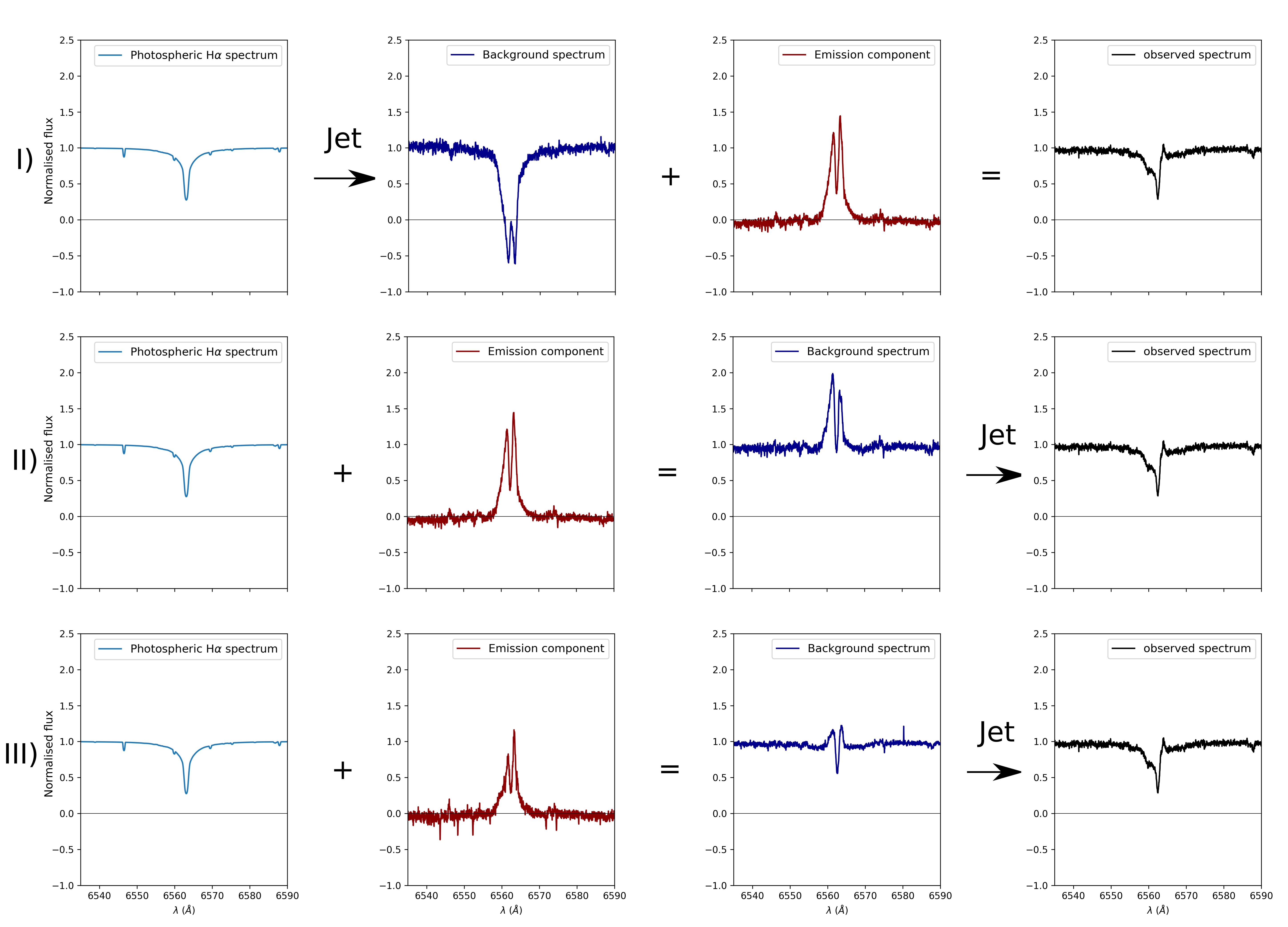}
\caption{Scenarios to determine the contribution of the emission component to the spectra. \textit{Upper row}: Case I, only the photospheric H$\alpha$ spectrum from the primary is absorbed by the jet (denoted by '${\scriptscriptstyle\xrightarrow{\text{jet}}}$'). The double-peaked emission component is constant and independent and is thus added (denoted by '+') to the resulting spectrum. \textit{Middle row}: Case II, the background spectrum is the combined photospheric H$\alpha$ line plus the emission component. This light is then absorbed by the jet. \textit{Lower row}: Case III, This is the same as case II, except for the fact that the emission component varies as a function of orbital phase. The emission component shown here is taken at phase $\phi=0.5$, and thus weaker than the emission component at phase $\phi=0$ (as can also be seen in Fig. \ref{fig:initialspec}).}\label{fig:backgroundspectra}
\end{figure*}
To test this fitting methodology for jets launched by post-AGB binaries, we start with the system \iras. the data for this object are part of a large radial velocity monitoring programme of post-AGB binaries by \cite{vanwinckel09}, which started in 2009 with the HERMES spectrograph, mounted on the 1.2m MERCATOR telescope, La Palma, Spain \citep{raskin11}. We choose post-AGB binary IRAS19135+3937 as test object for our fitting code, because we have a well-sampled orbital cycle for this object. Since the start of the monitoring campaign, we have obtained 90 spectra for this system, providing us with a good phase coverage of the orbital period. This post-AGB binary has a relatively short orbital period of $126.97\pm0.08\,$d and an eccentricity of $0.13\pm0.03$ \citep{gorlova15, oomen18}. The orbital parameters of the system are listed in Table \ref{tab:orbpar}. 
Ideally, we would be able to derive the radius of the post-AGB star by combining the distance to the object with its photometric flux. However, in the case of \iras, the complications are twofold. First, the distance from GAIA DR2 \citep{gaia16, gaia18} for \iras\, is an approximate estimate of the true distance, because the parallaxes in GAIA DR2 are determined using single-star models. Additionally, its orbital period ($\sim$\,127\,days, see Table~\ref{tab:orbpar}) implies that the GAIA DR2 parallax for this star will have a similar amplitude as its projected orbit, thereby reducing the accuracy of the resulting distance estimate. Second, this object is a semi-regular variable, which implies that its flux shows significant variability. This makes it impossible to accurately determine the stellar radius of the post-AGB star. Therefore, for the modelling of this system, we keep the stellar radius as a free parameter between set boundaries. 

The boundaries for the stellar radius are constrained based on its Roche-lobe filling factor\footnote{The Roche-lobe filling factor can be defined in terms of the potential of an equipotential surface relative to the potential of the surfaces through the Lagrange points L1 and L2 \citep{mochnacki84}. a Roche-lobe filling factor of 1 corresponds to surfaces inside the Roche lobe.}. A star that nearly fills its Roche lobe would be tidally distorted, causing ellipsoidal variations in its light curve \citep{wislon76}, which we do not observe for \iras. This implies that the radius of the post-AGB star must be smaller than its Roche lobe. In order to set an upper-limit for this parameter, we look at the Roche-lobe filling factor of sequence E red giants, which are binary systems that show ellipsoidal variations in their photometric data. From the sample of sequence E red giants in the study of \cite{nicholls10}, we derived an average Roche-lobe filling factor of $0.82\,\pm\,0.12\,R_\mathrm{RL}$. Based on these data, we set the upper boundary for the radius of the post-AGB star at 70\% of the Roche radius.

Fig. \ref{fig:dynspec} shows the dynamic spectra of IRAS19135+3937 made by 22 out of the 90 spectra. We choose to use these 22 spectra since they were taken during one cycle of 129.98\,d. The dynamic spectra are created by showing the observed H$\alpha$ line as a function of orbital phase, where we interpolated along the orbital phase for regions where we do not have observations. The absorption feature caused by the scattering of the evolved star's light in the jet is clearly visible between orbital phases $\theta = 0.3-0.7$. The depth, extent and duration of this absorption feature is dependent on the orbit of the binary system, the size of the evolved star, and the size, density, and velocity of the jet.


\subsection{Determination of the unobstructed H$\alpha$ spectral profile}
\label{sec:inputspec}
\begin{figure*}[h!]
\captionsetup{width=1.\textwidth}
\centering
  \begin{tabular}{@{}c@{}c}
  \includegraphics[width =.5\textwidth]{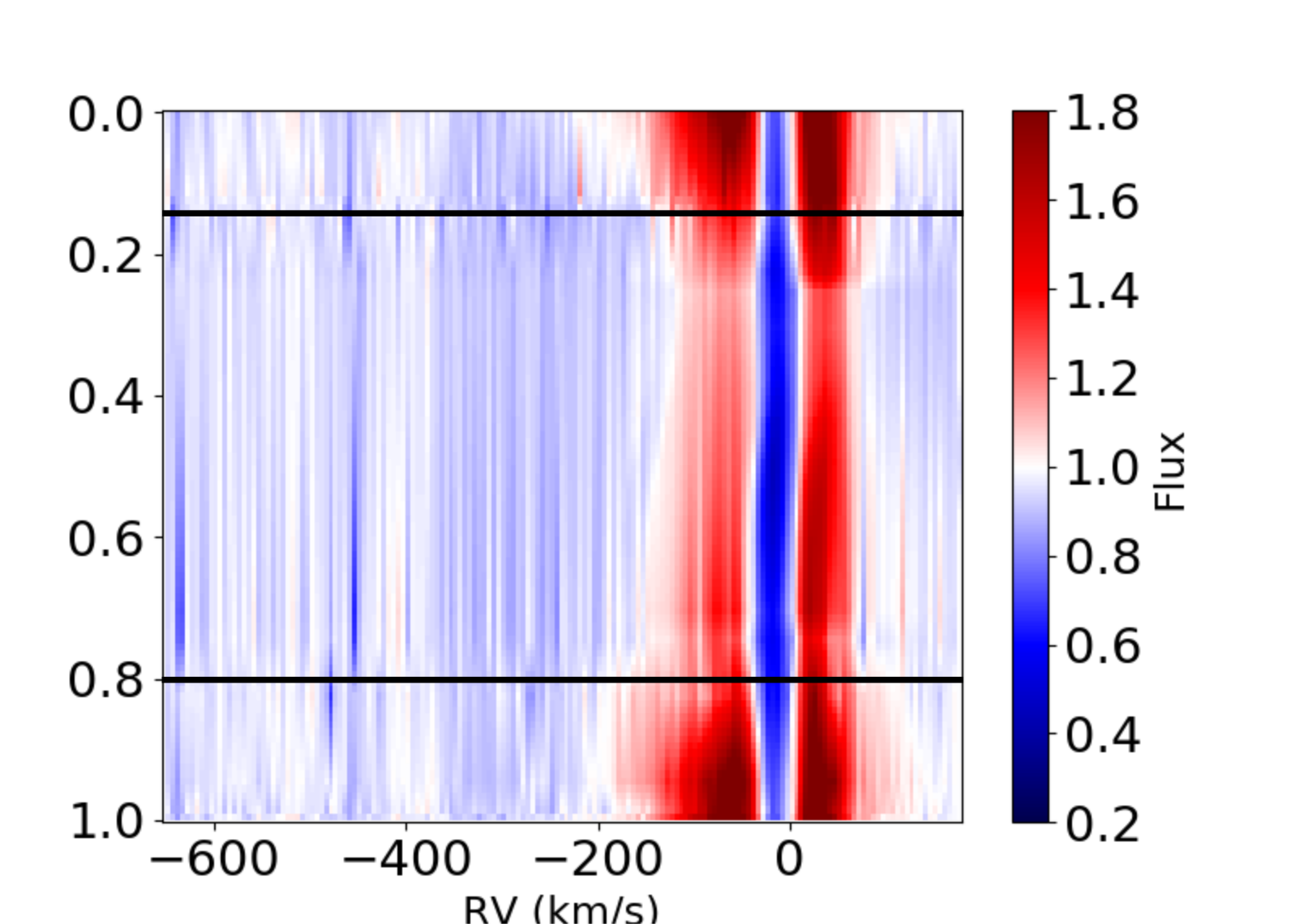}
&
	\includegraphics[width = .5\textwidth]{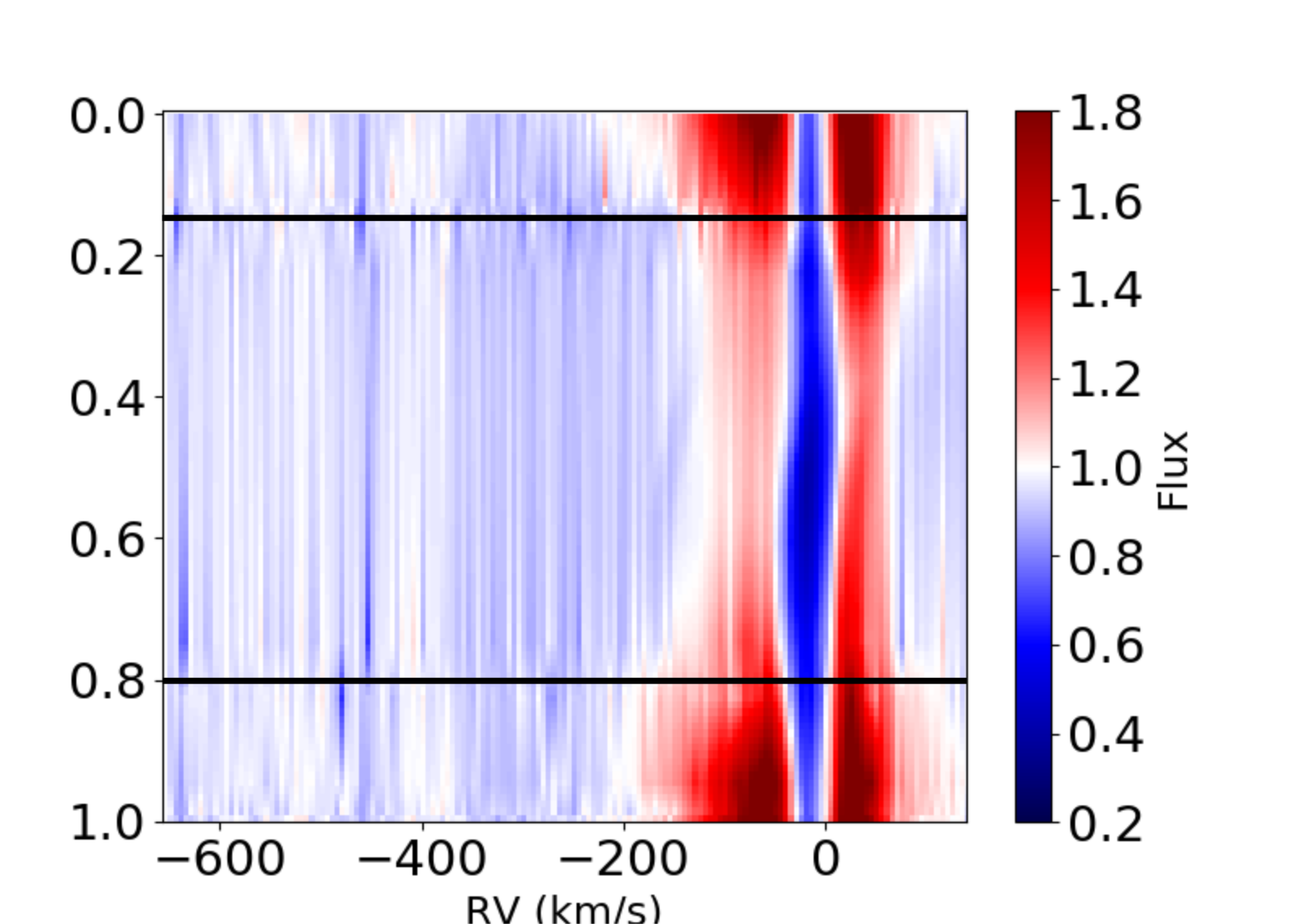}
\end{tabular}
\caption{Example of the determination of the spectra to be used as background light absorbed by the jet (case III). The regions between orbital phase $\phi\in[0.15-0.8]$ are substituted by values interpolated between adjacent spectra using a linear interpolation (\textit{left}) or a cubic interpolation (\textit{right}).}\label{fig:initialspec}
\end{figure*}
Our model calculates the amount of absorption of background light by the jet at each wavelength and returns the resulting synthetic spectrum. Hence, we have to provide our model with an initial background spectrum that does not include any absorption by the jet. Determining this initial H$\alpha$ spectrum is not straightforward, because the background spectrum is more complex than just the photospheric H$\alpha$ profile. 

The first component is the photospheric H$\alpha$ absorption line from the post-AGB stellar spectrum. A synthetic stellar spectrum from \cite{coelho05} with stellar parameters $T_\mathrm{eff} = 6000\,\mathrm{K}$, $\log\,g = 1.0$, and $[M/H] = -1.0$ was used to replicate this photospheric component \citep{gorlova15}. The absorption component is shifted at each orbital phase, according to the RV of the primary component.

In addition to the photospheric absorption, we observe an emission feature that is present throughout the whole orbital phase and is superimposed on the photospheric absorption from the AGB star. This feature varies between a double-peaked emission profile during inferior conjunction and a single, red-shifted emission peak during superior conjunction. During superior conjunction, the blue-shifted emission peak disappears at least in part due to the absorption caused by the jet. The emission component is not centred on the radial velocity of the post-AGB star nor that of the companion (see Fig. \ref{fig:dynspec}) so it is unclear if the emission is from the circum-companion accretion disk, from the base of the jet itself, or from yet another source.

If the emitting source is located behind the jet during the whole orbital phase, it must be included in the initial background spectra that we use to calculate the absorption. If the emitting source is not located behind the jet, on the other hand, the double-peaked emission profile should not be included in the background spectra, but should rather be added up to the resulting spectra after the absorption by the jet is determined. In addition, we need to determine if this emission feature is constant in strength or if it varies throughout the orbital period. 

We tested three possible situations, illustrated in Fig.~\ref{fig:backgroundspectra}: the emitting source is never blocked by the jet and is constant, the emitting source is blocked by the jet during superior conjunction and is constant, and the emitting source is blocked by the jet during superior conjunction and varies throughout the orbital phase.\\

\emph{Case I: The emitting source is not blocked by the jet and is constant in strength:} There are three components in each observed spectra: the photospheric absorption line, the double-peaked emission feature, and the absorption by the jet. The emission feature remains constant in this scenario, since it is not blocked by the jet. As shown in Fig. \ref{fig:backgroundspectra}, the background spectrum is the photospheric H$\alpha$ line of the post-AGB star. The absorption by the jet will thus only absorb light from this spectrum. Since the emitting source is not blocked by the jet, the resulting observed spectrum will be the sum of the absorbed spectrum and the emission component.

However, when we look at the absorbed spectrum for case I in Fig. \ref{fig:backgroundspectra}, we notice that the flux goes below zero at certain wavelengths. This is not possible, since the only source of light absorbed by the jet is the primary's photospheric light, the jet cannot absorb more than this light coming from the photospheric line. Hence, this scenario where the emission feature is not blocked by the jet is unphysical and we conclude that the emission region must be completely or partially behind the jet during superior conjunction. \\

\emph{Case II: The emitting source is blocked by the jet and constant in strength:} In this scenario, we assume the double-peaked emission feature remains constant in flux, but some of its light is absorbed by the jet. The variation in strength that we observe in the dynamic spectra of Fig. \ref{fig:dynspec} is thus only caused by the absorption by the jet. We tested this scenario by combining the photospheric H$\alpha$ line with the strongest double-peaked emission component that we observed in the spectra, as is shown in the middle row of Fig. \ref{fig:backgroundspectra}. The fitting routine will then calculate the amount of absorption by the jet and fit it to the data. This gives a very poor fit to our observations. The best-fitting jet solution is a jet that has an extremely high optical depth to account for the high amount of absorption needed in this case, which is not realistic since the jet is optically thin. We conclude that we are not able to obtain good fits with this set-up.\\

\emph{Case III: The emitting source is blocked or partially blocked by the jet and varies in strength:} We assume an intermediate situation, where the observed intensity of the emitting source varies throughout the orbital period. In order to determine the background spectrum at every orbital phase, we decided to interpolate the region between orbital phase $\phi\in[0.15-0.8]$ in the dynamic spectra, as is shown in Fig.\ref{fig:initialspec}. In this way, we assume that the background spectrum varies smoothly between phase 0.15, when the emitting source is presumably unobstructed, and phase 0.8, when the jet has completed its passage in front of the source. Since we do not know exactly how much emission the spectra show between these phases, we used two methods to interpolate this region: a linear interpolation (left plot of Fig. \ref{fig:initialspec}) and a cubic interpolation (right plot of Fig. \ref{fig:initialspec}). Both of these phase dependent spectral series are used in our model-fitting code as background spectra from which we subtract photons.

We include an extra uncertainty from the determination of the background spectrum, $\sigma_\mathrm{bs}$, in the total uncertainty used in Eq.~\ref{eq:logl}. This uncertainty is related to the strength of the emission component at each wavelength bin. The uncertainty in the flux will be larger for wavelength bins where this emission component is stronger. We quantify this by using the difference between the strongest observed emission flux and the flux of the background spectrum in case III as an extra uncertainty, as is shown in Fig.~\ref{fig:backgroundspectra}. The bottom panel of Fig.~\ref{fig:backgroundspectra} shows how the uncertainty is larger for wavelength bins with a stronger emission flux.

\begin{figure}
\centering
  \includegraphics[width =.5\textwidth]{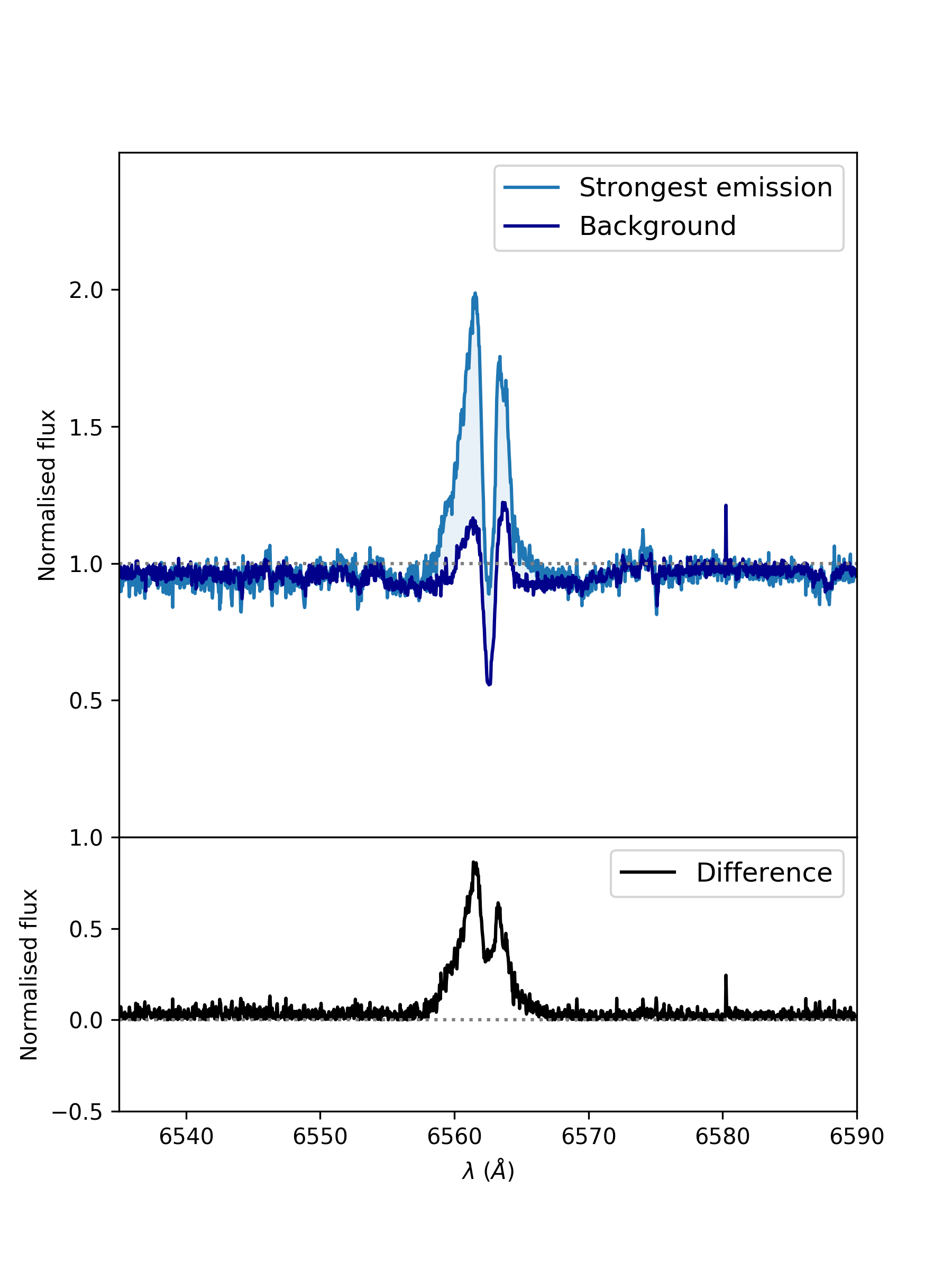}
\caption{The strongest emission spectrum for case II and the background spectrum used for case III (\textit{upper}). The uncertainty on the determination of the background flux is the difference between the flux of the strongest emission spectrum and the flux of the spectrum used for case III, where the emission varies throughout orbital phase (\textit{lower}).
}\label{fig:errorbackground}
\end{figure}


\section{Spatio-kinematic modelling of the jet}\label{sec:spatiomod}

We recreate the time-series of the H$\alpha$ spectral profile for \iras\, using the three jet configurations described in Sect. \ref{sec:model} in our model, in order to determine the spatio-kinematic structure of the jet. Several fits are performed for different values of the factor $p$ in the angle-dependent density profile in the stellar jet model, and the factors $p_\mathrm{in}$ and $p_\mathrm{out}$ in the X-wind and disk-wind model. These combinations are shown in Table \ref{tab:pval}. We created synthetic spectra that are used as background spectra for our fitting (see Sect. \ref{sec:inputspec}). The synthetic H$\alpha$ spectra contain the photospheric absorption line from the evolved star plus the double-peaked emission feature interpolated between orbital phase $\phi\in[0.15-0.8]$ using a linear and a cubic interpolation, as described in Sect \ref{sec:inputspec}. We eventually used these two dynamic background spectral series for our model fitting. By comparing the goodness of fit between the models that use the linearly interpolated background spectra and those using background spectra obtained by cubic interpolation, it is clear that the latter give better results for all runs. Hence, we will only focus on the models that use the background spectra obtained by cubic interpolation. In the following, we will further discuss our best-fitting synthetic spectra for the three jet configurations. 


\subsection{The best-fitting jet model}\label{sec:submodel}
\begin{figure*}[h!]
\captionsetup{width=1.\textwidth}
\centering
  \begin{tabular}{@{}c}
  \includegraphics[width=1\textwidth]{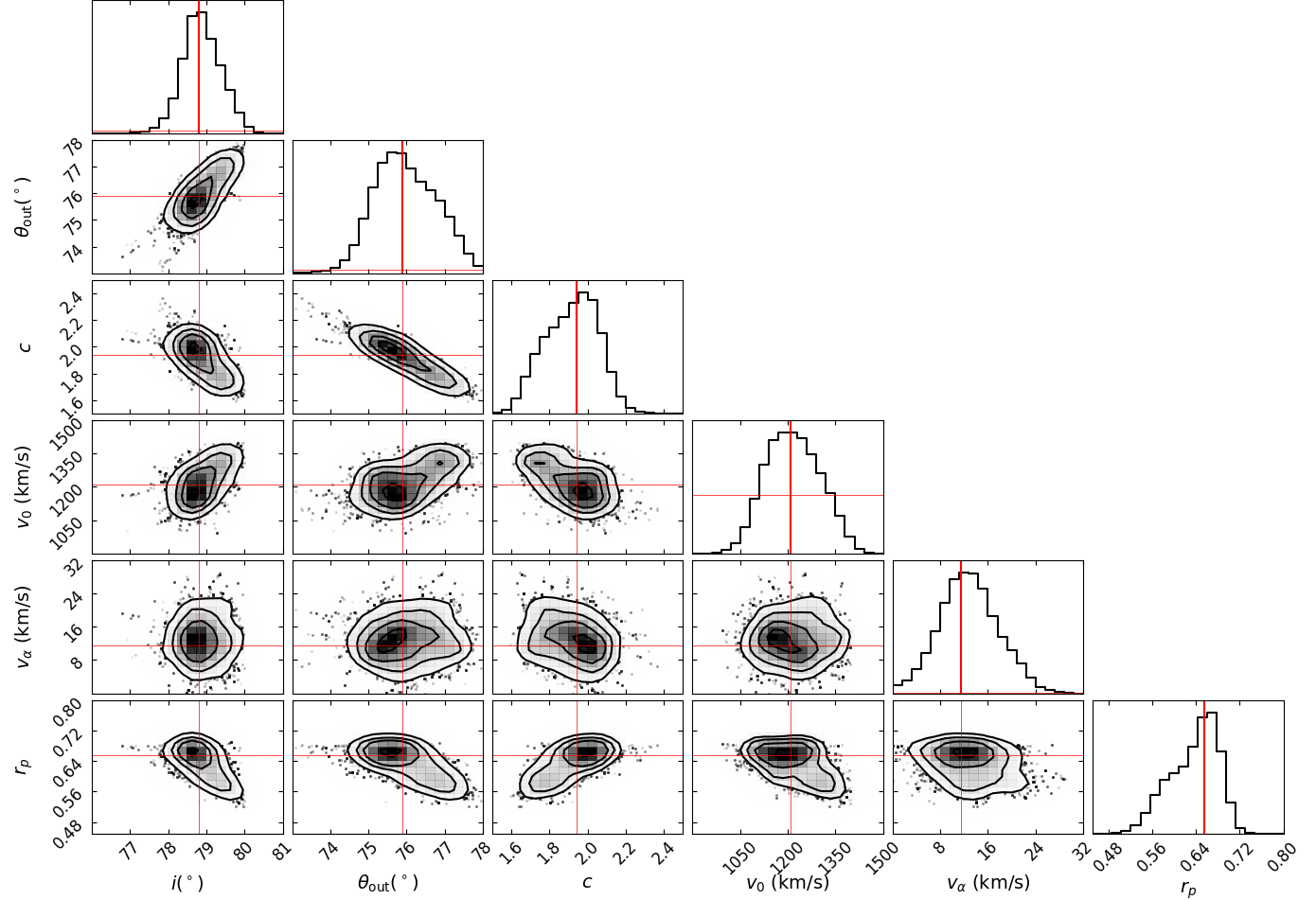}
\end{tabular}
\caption{Visualisation of the one- and two-dimensional projection of the posterior density distribution of our model parameters for the best-fitting model using the \texttt{corner} software by \cite{corner}. The red lines indicate the values for the best-fitting parameters (see Table \ref{tab:boundaries}).}\label{fig:corner}
\end{figure*}
\begin{figure}[h!]
\centering
\includegraphics[width=.42\textwidth]{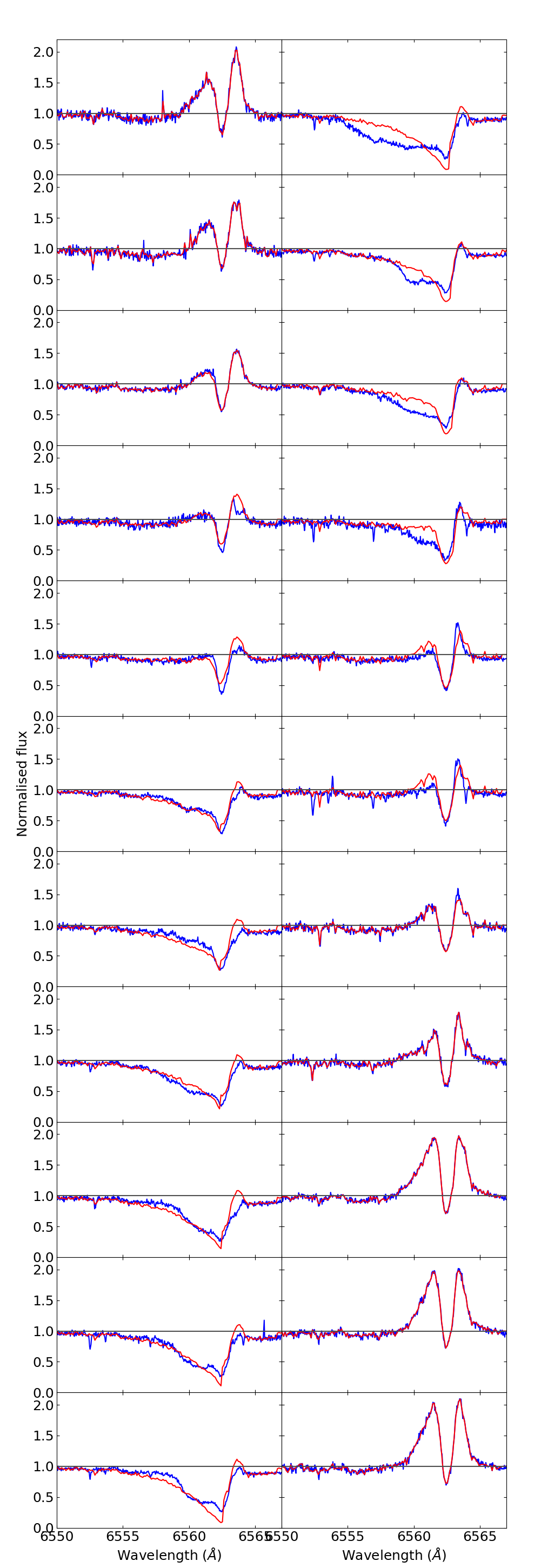}
\caption{Observations (\textit{blue}) vs. best-fitting model spectra for model A (\textit{red}). The panels increase in orbital phase from top to bottom and left to right.}\label{fig:allspec}
\end{figure} 
\begin{figure*}[h!]
\captionsetup{width=1.\textwidth}
\centering
  \begin{tabular}{@{}c}
  \includegraphics[width=1\textwidth]{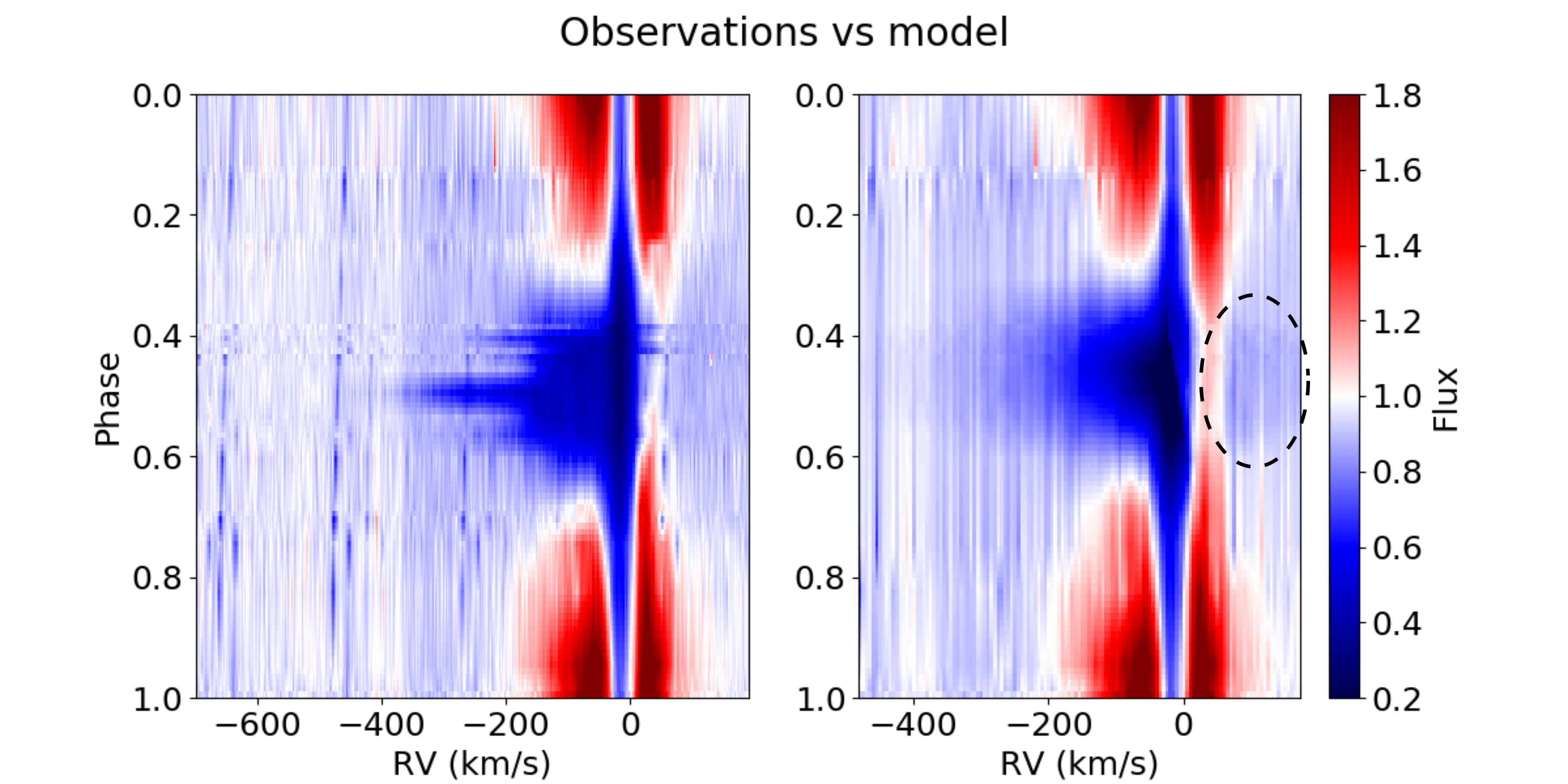}
\end{tabular}
\caption{Dynamic spectra created from the observed spectra (\textit{left}) and the model spectra (\textit{right}). The model dynamic spectra where created using the best-fitting model with the stellar jet configuration. The encircled region in the model spectra highlights the extra red-shifted absorption feature that is created by the model, but which we do not observe, as explained in Sect. \ref{sec:submodel}.}\label{fig:dynspecovsmod}
\end{figure*}
All results of our MCMC model-fitting routine are listed in Table \ref{tab:fitresults}, including their density structure, best-fitting parameters, and goodness of fit (reduced $\chi^2$ and Bayesian information criterion). The values of the model parameters are determined from the posterior density distribution of our MCMC-fitting routine. In order to extract these values and their corresponding standard deviations, we fit a Gaussian curve to the posterior density distribution. The mean and standard deviation of this Gaussian curve are reported alongside the best fit values in Table~\ref{tab:fitresults}.

Since we are comparing the fitting results of different configurations with a different set of parameters, we evaluate the relative goodness of fit of each model by comparing their Bayesian information criterion (BIC), which is defined as:
\begin{equation}
    \mathrm{BIC} = \ln(n)k - 2\ln\mathcal{L},
\end{equation}
with $n$ the number of data points and $k$ the number of model parameters. The BIC takes into account the number of parameters in a model. Hence a higher number of parameters gets penalised by a higher BIC value. 

The best-fitting model is the model using the stellar jet configuration with the background spectrum obtained by cubic interpolation and p-factor $p=8$ (see Table \ref{tab:boundaries} for corresponding best-fitting parameters). The posterior density distribution of the model is represented in the corner plot of Fig. \ref{fig:corner}. The individual spectra and dynamic spectra of this model are shown in Figs. \ref{fig:allspec} and \ref{fig:dynspecovsmod}, respectively. The jet has a very wide half-opening angle of $\theta_\mathrm{in} = 76\degr$. 

The jet for this model reaches a maximum velocity of $v_0 = 1210\,$km$\,s^{-1}$ along the jet axis. Due to the large inclination angle of $i=79\degr$, the projected jet velocities reach blue-shifted values of $\approx\,400\,$km$\,s^{-1}$, as is the case for the observations. In order to determine the nature of the companion, we could compare these jet velocities with typical escape velocities of stellar objects. For a $1M_\odot$ white dwarf, the escape velocity is about $6500\,$km s$^{-1}$. For a $1M_\odot$ main sequence star, this is about $618\,$km s$^{-1}$. Since the observed jet velocities are closer to the escape velocity of a main sequence star, we conclude that the companion in this system is most likely a main sequence star. This conclusion is also supported by \cite{oomen18}, who found that based on the initial mass of the companion it is likely a $1M_\odot$ main sequence star.
 
The absorption feature caused by the jet covers a large part of the orbital phase (between $\phi\in[0.25-0.75])$. This suggests that the difference between orbital inclination and jet half-opening angle should be rather small. Since the absorption is not seen throughout the whole orbital phase, the jet angle is most likely smaller than the inclination angle. This is also the case in the results of our best-fitting model, where the jet has a half-opening angle of $\theta_\mathrm{out}=76\degr$ and the inclination angle is slightly larger ($i=79\degr$).
\begin{figure}
\centering
\includegraphics[width=.5\textwidth]{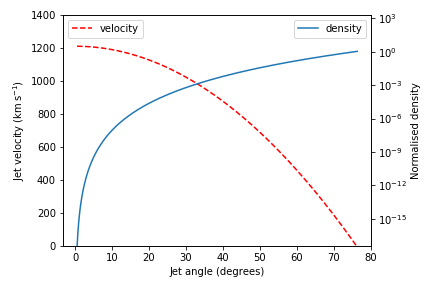}
\caption{Velocity and density structure as a function of jet angle for the best-fitting model. The density is normalised such that $\rho_\mathrm{edge} = 1$.}\label{fig:bestfit}
\end{figure}
\begin{figure}[h!]
\centering
  \includegraphics[width=.5\textwidth]{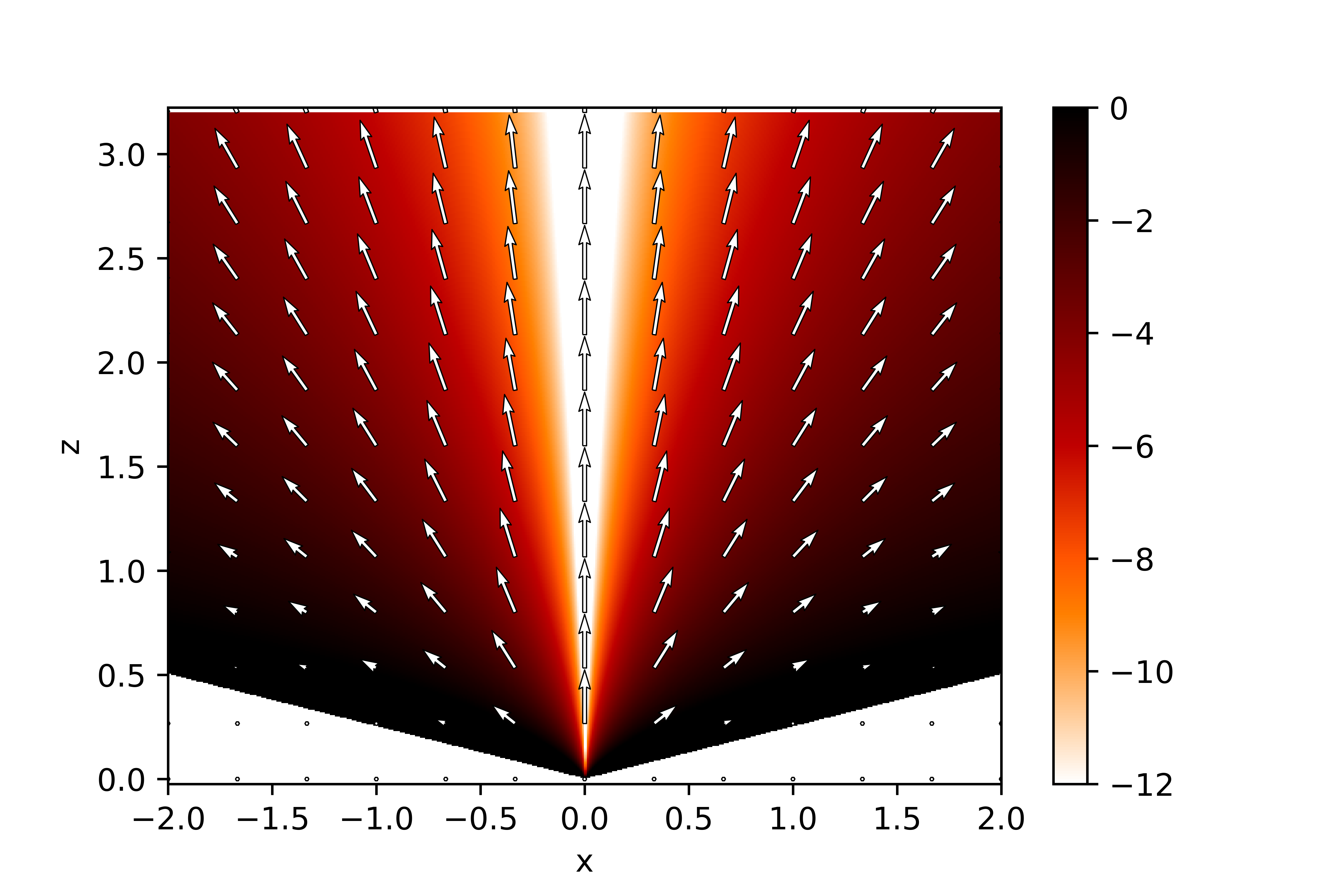}
\caption{Velocity field and density structure for the best-fitting model of the jet in \iras. The colour of the density structure is represented in a logarithmic scale.}\label{fig:veldens}
\end{figure}
Figs. \ref{fig:allspec} and \ref{fig:dynspecovsmod} show a fairly good match between the model spectra and the observations. An interesting feature that appears in the model spectra, but not in the observed spectra is the red-shifted absorption feature, indicated in Fig. \ref{fig:dynspecovsmod} by the dashed circle. This absorption feature in the model is due to absorption by the jet in the region where the light rays enter the jet. The jet particles in this part of the jet have a positive velocity when projected on to the line-of-sight through the jet. Hence, the absorption will be red-shifted. This absorption is not as significant in the observations as in the model calculations.

The velocity and density structure of this model, represented in Figs. \ref{fig:bestfit} and \ref{fig:veldens}, show that the jet is very dense at the edges and has an extremely low density along the jet axis. 


\subsection{Comparison of the three jet models}
\begin{figure}
\centering
\includegraphics[width=.5\textwidth]{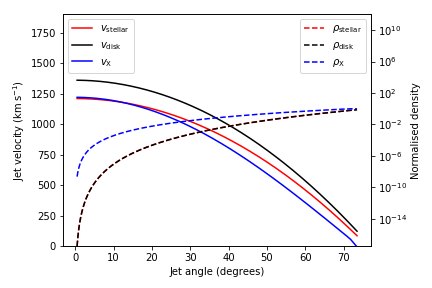}
\caption{Velocity and density structure as a function of jet angle for the best-fitting model for the stellar jet model (A), the disk wind model (B), and the X-wind model (C). The density is normalised such that $\rho_\mathrm{edge} = 1$.}\label{fig:rhovbestall}
\end{figure}
Here, we investigate which of the three models gives the best representation of the jets in our post-AGB binary systems. In order to differentiate which model fits the data best, we need to look at the difference between the BIC values of the models, where a lower BIC value translates into a better fit. We find the goodness of fit for the best-fitting model of the stellar jet (model A), the disk wind (model B), and the X-wind (model C) to be BIC$_\mathrm{A}=-24878$, BIC$_\mathrm{B}=-24867$, and BIC$_\mathrm{C}=-24805$, respectively. According to  \cite{kass95}, a difference of $|$BIC$_1 - $BIC$_2| > 6$ is a strong condition to determine that the model with the lowest BIC value fits the data better. This only works when the absolute BIC values are completely correct. In our case, the relative BIC values are reliable, but the absolute BIC values are not. The BIC value is highly dependent on the errors in the $\chi^2$ calculations. Our error only takes into account the error on the spectra (determined from the signal-to-noise ratio) and the uncertainty from the determination of the initial background spectra, which is given as input to the model.

Thus, in order to compare the different models, we will compare both their BIC values and the reduced $\chi^2$ values in order to determine the best-fitting models. By doing so, it can be seen that model A fits best to the observations. This model only gives a slightly better fitting result than models B \& C, however. The reduced $\chi^2$ for model B is even slightly better than that of model A ($\chi^2_{red,B} = 0.2428$ and $\chi^2_{red,A} = 0.2434$). If we would neglect that model B has more model parameters than model A (8 compared to 6), this would be the better fitting model. However, since we do want to penalise for extra model parameters, we conclude that model A has a slightly better fit than model B ($\Delta$BIC$=11$). 

The BIC of the best-fitting X-wind configuration (model C) is higher than model A with $\Delta $BIC$=73$. For this model, the model parameters are also very similar to model A. The main difference between model C and model A is a smaller inclination angle and jet angle. These two angles are smaller by about 3$\degr$ for model C. 

In Fig. \ref{fig:rhovbestall}, we show the velocity profile and density profile in the jet for best-fitting models A, B, \& C. This shows again that all three models tend to converge to a similar structure. The inner jet boundary of model B \& C in particular tends to lie very close to the outer jet edge. Hence, for these model results, the jet is mainly governed by the inner velocity and density law. The ``outer jet" region is a relatively small, negligible region in these models.

The half-opening angle of the jet lies between $72\degr$ and $77\degr$ for all three models. The velocity along the edges of the jet is about $10-20\,$km s$^{-1}$ and reaches maximal velocities of $950-1350\,$km s$^{-1}$ at the centre. The best fitting density law is the same for models A \& B. Both model B and model C converge towards a similar simple structure as the stellar jet of model A. We expect that for some post-AGB binary systems, such as \iras, the three models will converge to a similar structure, whilst for other systems, the three models will be easier to differentiate results. This is due to the large diversity in terms of orbital parameters and system sizes in these systems.

The best-fitting model for the stellar jet configuration has a density parameter $p=8$. This configuration assumes a simple conical outflow with an increasing density from the pole towards the jet edge. The best-fitting model for the disk wind configurations has a similar density law in the inner jet ($p_\mathrm{in}=8$). The outer density law in model B and model C is $p_\mathrm{out}=-2$. The inclination for all three configurations converges to $\sim 75-80\degr$, which implies that the system is observed nearly edge-on. The study by \cite{gorlova15} has suggested this as well. Their conclusion was based on the semi-regular variability of the system, first observed by \cite{sallman04}. \cite{gorlova15} showed that the light variations are most likely caused by the partial obscuration of the evolved star by the puffed-up circumbinary disk during inferior conjunction. This obscuration can only occur if the inclination of the system is rather high, which is what we find.

Similarly to the inclination, the jet velocity at the edge and the jet angle for each of the three configurations converges to the same values. This means that the density and velocity structure in the different configurations will be alike as well, as can be seen in Fig. \ref{fig:rhovbestall}. 

By knowing both the orbital parameters and the inclination of the orbital system, we can calculate the radius of the evolved star and the mass of the companion. Given the mass function and the inclination found in the best-fitting model ($i=79\degr$), we calculate the mass of the companion star to be $0.44$\,M$_\odot$. The exact radius of the evolved star was a model parameter in our code, given in units of the semi-major axis of the primary. The result for radius of the evolved star in the best-fitting model is $0.66\,\pm\,0.04\cdot a_\mathrm{prim}$, with $a_\mathrm{prim}$ the semi-major axis. Hence, given the projected semi-major axis $a_\mathrm{prim}\cdot\sin\,i$ (Table \ref{tab:orbpar}) and the inclination found from our model-fitting routine, the radius of the evolved star will be R$_\mathrm{prim}\,=30\,\pm\,3\,$R$_\odot$.


\subsection{A possible cavity in the jet}

We find that the density in the inner part of the jet is extremely low, compared to the outer part of the jet (see Fig.\ref{fig:bestfit}). This leads us to question whether the inner region contributes at all to the absorption caused by the jet. We test this by including a cavity (region of zero density) in the best-fitting jet model. We increase the jet cavity from $0\degr$ up to $50\degr$, and compared the BIC values of these tests with the best-fitting model. By increasing the cavity to have opening angles from $0\degr$ to $30\degr$, the synthetic spectra, and thus the goodness of fit, does not change. The density in this region in the best-fitting model is just too low in order to contribute to the absorption feature in the spectra. By further increasing the cavity in the jet from $30\degr$, the BIC improves slightly and reaches the same value again when the cavity is increased up to $\sim\,40\degr$.

By including this cavity in our model, we show that the inner region of the jet ($<\,40\degr$) has extremely low density. In our best-fitting models for both the X-wind and the disk wind, the bulk of the matter is launched at an angle $>\,40\degr$. This is in accordance with the disk wind theory by \cite{blandford82}, where the magnetic field line and the jet axis must make an angle of $>\,30\degr$. This does not mean that the inner regions are completely void of matter, but rather that the density is too low to have any effect on the H$\alpha$ line profile. This consequently impacts the maximum outflow velocity in the jet. Assuming that the inner region does not contribute to the observed absorption feature, since it is too low a density, we can conclude that the maximum velocity in the jet is reached at an angle of about $40\degr$ (the innermost region of the jet that launches matter). For the best-fitting stellar jet model, this means that the jet reaches velocities of about 870~km\,s$^{-1}$, instead of 1210 km\,s$^{-1}$.

Interestingly, the existence of a cavity seems to be implied by the data. In future work, we will alter the density laws, such that the disk wind and X-wind models are closer to their initial conception launching gas in a cone geometry without a central outflow.

\section{Summary and future work}\label{sec:conclusions}
We developed a fitting procedure that allows us to determine the spatial, velocity, and density structure of jet-creating post-AGB binaries. Our procedure implements three separated jet models: a stellar jet configuration (Model A), a disk wind configuration (Model B), and an X-wind configuration (Model C); which are based on jet launching mechanisms designed to explain YSO jets. Our procedure calculates synthetic spectral lines for these systems by calculating how much of the evolved star's light is scattered by the jet as it passes between the post-AGB star and the observer. These synthetic spectral lines were then fitted to the observations. This shows that the companion is likely a main sequence star.

We fitted synthetic lines to the data of the post-AGB binary \iras\, for each of the three jet configurations. The best-fitting model is the stellar jet configuration, where the jet is governed by the density law: $\rho\propto\theta^{8}$. The jet in this model is a very wide outflow of $76\,\degr$, with a slow velocity component along the edges ($v_\mathrm{out}= 11\,$km\,s$^{-1}$) and a high-velocity component in the inner regions. The inner region in the jet has extremely low density, making the contribution of this region to the scattering in our model insignificant. The highest velocity in the jet is about $870\,$km\,s$^{-1}$. There is no significant difference between the results of the three jet configurations. Both the X-wind and the disk wind model converge to a solution similar to the stellar jet configuration.  We conclude that our model succeeds in fitting the dynamic spectra well. 

Unexpectedly, we detected an emission component that is of unknown origin. Our fits indicate that this source varies over the orbit and that this light is absorbed when the companion and its jet are at inferior conjunction.

In future work, we will apply our fitting procedure to our target sample of about 15, jet-creating, post-AGB binaries. We will apply this generic model to determine the configurations of all objects and study if we can infer additional information on the jet launching mechanisms for these systems. We also expect to be able to measure the actual jet density more accurately, something that will provide us with the accretion rates. Ultimately, we would like to be able to connect accretion rates, AGB star abundances, and circumbinary disk longevity in one narrative that will shed light on the origin of post-AGB binaries.

\begin{acknowledgements}
This work has made use of data from the European Space Agency (ESA) mission
{\it Gaia} (\url{https://www.cosmos.esa.int/gaia}), processed by the {\it Gaia}
Data Processing and Analysis Consortium (DPAC,
\url{https://www.cosmos.esa.int/web/gaia/dpac/consortium}). Funding for the DPAC
has been provided by national institutions, in particular the institutions
participating in the {\it Gaia} Multilateral Agreement. This work was performed on the OzSTAR national facility at Swinburne University of Technology. OzSTAR is funded by Swinburne University of Technology and the National Collaborative Research Infrastructure Strategy (NCRIS). DK acknowledges the support of the Macquarie University New Staff funding. HVW acknowledges support from the Research Council of the KU Leuven under grant number C14/17/082.  Based on observations obtained with the HERMES spectrograph, which is supported by the Research Foundation - Flanders (FWO), Belgium, the Research Council of KU Leuven, Belgium, the Fonds National de la Recherche Scientifique (F.R.S.-FNRS), Belgium, the Royal Observatory of Belgium, the Observatoire de Gen\`eve, Switzerland and the Th\"uringer Landessternwarte Tautenburg, Germany.

\end{acknowledgements}


\bibliographystyle{aa}
\bibliography{allreferences.bib}

\newpage
\begin{appendix}
\section{Table with all fitting results for the three models.}
In Table~\ref{tab:fitresults} we present the best fitting parameters with corresponding reduced chi-squared and BIC for the fitting of the three jet models to IRAS19135+3937. $\Delta$BIC denotes the difference of the BIC with the BIC of the best-fitting model. The jet models are model A (the stellar jet model), model B (the disk-wind model), and model C (the X-wind model).
\newpage
\begin{sidewaystable}
\setlength\tabcolsep{4pt}  
    \centering
\longtab{1}{
\begin{longtable}{clllllllllllll}
\caption{Best fitting parameters with corresponding reduced chi-squared, log-likelyhood, and BIC for the fitting of the three jet models to IRAS19135+3937. $\Delta$BIC denotes the difference in BIC with the best-fitting model. The jet models are model A (the stellar jet model), model B (the disk-wind model), and model C (the X-wind model).}\label{tab:fitresults} \\
\hline
\hline \\ 
Jet model & $p$\footnote{$p$ for model A, $p_\mathrm{in}$ and  $p_\mathrm{out}$ for model B \& C.} & $i$  & $\theta_\mathrm{out}$ &  $\theta_\mathrm{in}$ & $c$ & $v_0$ & $v_\mathrm{in}$ & $v_\mathrm{out}$ & $c_v$ & $R_\mathrm{prim}$ & $\chi^2$& BIC & $\Delta$BIC \\
& & degrees & degrees & degrees & & km\,s$^{-1}$ & km\,s$^{-1}$ & km\,s$^{-1}$ &  & a$_\mathrm{prim}$ \\
 \\ \hline  \\
\endfirsthead
\endhead
\hline
\endfoot
A  & 8 & $78.8\pm0.8$ & $75.9\pm1.1$ &  & $1.94\pm0.14$ & $1210\pm90$ &  & $11\pm5$ &  & $0.66\pm0.04$ & 0.2434 & -24878 & 0 \\
 & 4 & $78.5\pm1.1$ & $76.3\pm0.9$ &  & $2.16\pm0.08$ & $1010\pm40$ &  & $16\pm4$ &  & $0.68\pm0.02$ & 0.2461 & -24844 & 34 \\
 & 2 & $79.9\pm1.1$ & $77.8\pm1.5$ &  & $2.11\pm0.12$ & $1150\pm170$ &  & $18\pm4$ &  & $0.56\pm0.03$ & 0.2545 & -24738 & 140 \\
 
\\ \hline \\

B & 8/2 & $79.5\pm0.9$ & $76.3\pm0.8$ & $76.2\pm0.8$  & $2.83\pm0.04$ & $1360\pm30$ & & $82\pm6$ & $0.22\pm0.07$ & $0.565\pm0.011$ & 0.2433 & -24861 & 17\\
 & 8/-2 & $79.5\pm0.8$ & $76.4\pm0.9$ & $76.3\pm1.5$  & $2.82\pm0.13$ & $1360\pm30$ & & $77\pm9$ & $0.30\pm0.12$ & $0.559\pm0.014$ & 0.2428 & -24867 & 11\\
 & 4/2 & $79.6\pm0.8$ & $77.2\pm0.7$ & $77.0\pm0.8$  & $2.46\pm0.04$ & $1240\pm40$ & & $70\pm4$ & $0.61\pm0.06$ & $0.525\pm0.013$ & 0.2463 & -24824 & 54\\ 
 & 4/-2 & $79.6\pm0.7$ & $77.3\pm0.9$ & $76.9\pm0.9$  & $2.47\pm0.04$ & $1240\pm30$ & & $69\pm4$ & $0.60\pm0.12$ & $0.52\pm0.02$ & 0.2455 & -24833 & 45\\
 & 2/-2 & $76.5\pm0.9$ & $75.1\pm0.9$ & $74.4\pm1.1$  & $2.39\pm0.06$ & $980\pm20$ & & $82\pm16$ & $0.47\pm0.10$ & $0.56\pm0.02$ & 0.2517 & -24755 & 123\\
  & 0/4 & $79.5\pm0.8$ & $77.6\pm1.3$ & $77.2\pm1.6$  & $2.05\pm0.08$ & $1360\pm100$ & & $67\pm9$ & $0.83\pm0.08$ & $0.47\pm0.06$ & 0.2600 & -24651 & 227\\
 \\\hline \\
 
 C  & 8/2 & $75.6\pm1.0$ & $72.4\pm1.3$ & $70\pm2$  & $2.96\pm0.06$ & $1300\pm60$ & $59\pm18$ & $5\pm4$ &   & $0.65\pm0.02$ & 0.2536 & -24731 & 147 \\
 & 8/-2 & $75.6\pm1.6$ & $72\pm3$ & $68\pm5$  & $2.98\pm0.16$ & $1280\pm80$ & $61\pm17$ & $5\pm5$ &   & $0.657\pm0.014$ & 0.2790& -24412 & 466 \\
 & 4/2 & $75.4\pm1.2$ & $73\pm2$ & $55\pm11$  & $2.4\pm0.2$ & $1340\pm50$ & $430\pm50$ & $19\pm4$ &   & $0.66\pm0.03$ & 0.2501 & -24776 & 102 \\
& 4/-2 & $75.4\pm0.7$ & $73.0\pm0.9$ & $71.7\pm1.4$  & $2.63\pm0.08$ & $1220\pm50$ & $60\pm20$ & $12\pm10$ &   & $0.662\pm0.013$ & 0.2477 &-24805 & 73 \\
  & 2/-2 & $76.2\pm0.9$ & $74.9\pm1.2$ & $62.5\pm1.0$  & $2.48\pm0.08$ & $1310\pm60$ & $300\pm20$ & $27\pm7$ &   & $0.56\pm0.03$ & 0.2544 & -24722 & 156 \\
  & 0/4 & $75.3\pm0.9$ & $73.0\pm1.1$ & $44\pm4$  & $1.8\pm0.2$ & $1260\pm40$ & $
  760\pm80$ & $20\pm5$ &   & $0.67\pm0.02$ & 0.2491 & -24788 & 90 \\

\end{longtable}
}
\end{sidewaystable}

\end{appendix}
\end{document}